\documentclass[twocolumn,preprintnumbers,amsmath,amssymb,prd,superscriptaddress,floatfix]{revtex4-1}
\pdfoutput=1
\usepackage{graphicx}
\graphicspath{{figures/}}
\usepackage{hyperref}
\usepackage{amsfonts}
\usepackage{mathrsfs}
\usepackage[utf8]{inputenc}
\usepackage[capitalise]{cleveref}
\usepackage{orcidlink}
\begin{document}
\preprint{RIKEN-iTHEMS-Report-26}
%%%%%%%%%%%%%%%%%%%%%%%%%%%%%%%%%%%%%%%%%%%%%%%%%%%%%%%%%%%%
%%%%%%%%%%%%%%%%%%%%%%%%%%%%%%%%%%%%%%%%%%%%%%%%%%%%%%%%%%%%

\title{Inverted Pendulum in the Sky: Floquet Stabilization of Cosmological Axions}

\author{Tokiro Numasawa\, \orcidlink{0000-0003-1024-7907}}
\affiliation{Institute for Solid State Physics, University of Tokyo, Kashiwa 277-8581, Japan}

\author{Masahito Yamazaki\,\orcidlink{0000-0001-7164-8187}}
\affiliation{Department of Physics, University of Tokyo, Tokyo 113-0033, Japan}
\affiliation{Kavli Institute for Physics and Mathematics of the Universe (Kavli IPMU), UTIAS, University of Tokyo, Chiba 277-8583, Japan}
\affiliation{Center for Data-Driven Discovery (CD3), Kavli IPMU, UTIAS, University of Tokyo, Kashiwa, Chiba 277-8583, Japan}
\affiliation{Trans-Scale Quantum Science Institute, University of Tokyo, Tokyo 113-0033, Japan}
\affiliation{Center for Interdisciplinary Theoretical and Mathematical Sciences (iTHEMS),\\ RIKEN, Saitama 351-0198, Japan}

\begin{abstract}
Many cosmological scenarios require a scalar field to sit near the top of its potential, which usually requires a fine-tuned initial condition.
We ask whether the hilltop of an axion potential can be dynamically stabilized by a rapidly oscillating coupling, as for the inverted pendulum of Kapitza, and to what extent this can reduce the fine-tuning.
Using Floquet theory, we determine the stability window of the hilltop and analyze several realizations: an oscillating coupling, a spinning or oscillating second field, and a gyroscopic analogue, an inverted spinning top realized by a sigma model on a squashed three-sphere.
In an expanding Universe the stabilized phase lasts for a number of $e$-folds logarithmic in the hierarchy of scales, independently of the cosmological background.
Classifying the stabilizers by how their effective frequency redshifts, we show that in these realizations the hilltop cannot both attract the field and leave it frozen by the Hubble friction during radiation or matter domination, and that during inflation the attraction is limited by the Hubble friction and the energy budget.
Thus the simple realizations studied here do not generically remove the hilltop fine-tuning.
They can instead delay the onset of the oscillation and enhance the relic abundance as $n_\phi/n_{\phi,{\rm std}}\propto(\tilde m/H_c)^{3/2}$ in radiation domination.
We discuss the implications for axion dark matter, inflation, quintessence and curvatons.
\end{abstract}

%%%%%%%%%%%%%%%%%%%%%%%%%%%%%%%%%%%%%%%%%%%%%%%%%%%%%%%%%%%%
\maketitle
%%%%%%%%%%%%%%%%%%%%%%%%%%%%%%%%%%%%%%%%%%%%%%%%%%%%%%%%%%%%
\section{Introduction}
%%%%%%%%%%%%%%%%%%%%%%%%%%%%%%%%%%%%%%%%%%%%%%%%%%%%%%%%%%%%

Scalar fields sitting close to a local maximum of their potential
appear in many different contexts in cosmology and at many different energy scales:
hilltop inflation \cite{Boubekeur:2005zm} (see also \cite{Freese:1990rb,Nomura:2017ehb,Nomura:2017zqj} for natural inflation, whose potential is that of an axion),
the axion misalignment mechanism \cite{Preskill:1982cy,Abbott:1982af,Dine:1982ah} with an initial angle near the hilltop
\cite{Turner:1985si,Lyth:1991ub,Bae:2008ue,Visinelli:2009zm,Kobayashi:2013nva,Co:2018mho,Takahashi:2019pqf},
hilltop quintessence \cite{Dutta:2008qn,Choi:1999xn,Ibe:2018ffn,Murayama:2018lie,Bhattacharya:2024kxp},
and the hilltop curvaton \cite{Kawasaki:2011pd,Fukuda:2018haz}.
Hilltops are also favored by the refined de Sitter swampland conjecture \cite{Garg:2018reu,Ooguri:2018wrx}, which can be satisfied by a sufficiently negative curvature of the potential \cite{Fukuda:2018haz}.
In all these scenarios one faces the same question: why does the
field stay near the hilltop?
As long as the Hubble parameter $H$ is larger than the (tachyonic) mass
of the field, the Hubble friction keeps the field frozen.
Once $H$ drops below the mass, however, the field rolls down the potential
within a few Hubble times, unless the initial condition is tuned exponentially close to the top.

In classical mechanics, there is a celebrated loophole to the statement
that a maximum of the potential is unstable:
a pendulum whose pivot vibrates rapidly in the vertical direction
can be stable in the inverted position.
This is the Kapitza pendulum \cite{Kapitza:1951,Landau:Mechanics}.
The rapid vibration generates an effective potential
for the slow motion, which turns the maximum into a local minimum
once the amplitude of the vibration exceeds a critical value.
The same mathematics underlies the Paul trap of charged particles \cite{Paul:1990zz}, and
is also a prototypical example of Floquet engineering
of time-periodic systems \cite{Floquet:1883,Goldman:2014xja,Bukov:2015dgu}.
Note also that the pendulum analogy has been fruitful in other contexts
of astroparticle physics, see e.g.\ \cite{Hannestad:2006nj}.

In this paper, we ask whether a similar mechanism can work for
cosmological scalar fields, and in particular for axions whose
potential is periodic, so that the analogy with the pendulum is literal.
We will see that the hilltop can indeed be dynamically stabilized when the axion couples to a rapidly oscillating background,
but that the expansion of the Universe severely limits what this can achieve.
Our aim is to clarify this situation systematically, rather than to advocate a particular mechanism.
The appropriate language is Floquet theory:
linearized fluctuations around the hilltop obey
a Mathieu-type equation, the same equation that governs
preheating after inflation \cite{Traschen:1990sw,Kofman:1994rk,Shtanov:1994ce,Kofman:1997yn} (see \cite{Amin:2014eta,Lozanov:2019jxc} for reviews).
While in preheating one is interested in the instability bands of the Mathieu equation,
here we are interested in the stability region that
extends to negative values of the Mathieu parameter $\mathsf{a}$,
which corresponds to a tachyonic mass.

The new ingredient in cosmology is the Hubble friction, which
damps not only the axion but also the agent of the oscillation itself.
Correspondingly, we will find that the stabilized phase
lasts only for a finite number of $e$-folds,
which is generically logarithmic in the hierarchy of scales involved.
More importantly, whether the hilltop can attract the field and leave it there is controlled by how the effective frequency of the stabilized oscillation redshifts compared with the Hubble parameter.
We classify the realizations in these terms (\cref{tab:classification}) and find that, in radiation or matter domination, none of them can bring the field towards the hilltop and leave it frozen by the Hubble friction,
while in de Sitter space the attraction is limited by the Hubble friction and the energy budget.
The stabilization can instead be used to delay the onset of the oscillation of the field, which is the most direct cosmological application of the mechanism that we find.

Let us comment on related works.
The closest is the recent work \cite{Murai:2025wbg}, which introduced a ``Kapitza-type modulation'' of a light axion by a heavy oscillating axion.
There the averaged potential of the light axion is multiplied by a Bessel function of the oscillation amplitude and flips its sign, so that the hilltop becomes a minimum;
in the language of the Hill equation \eqref{eq:Hill} below, this is the case where the averaged coefficient satisfies $V_0+\bar B<0$, an effect at leading order in the time average (i.e.\ in the inverse frequency).
The classic Kapitza effect, on which we focus, arises at second order, with the hilltop remaining a maximum of the averaged potential at first order (see \cref{sec:clean}).
The two mechanisms share several consequences, notably the delayed onset of the oscillation and the resulting enhancement of the relic abundance found in \cite{Murai:2025wbg}.
The central new result of this paper is a classification of the cosmological evolution by the redshift of the effective restoring frequency, which yields general bounds on the ability to attract a field to the hilltop and subsequently freeze it there.
We also determine the Floquet stability window and its duration, show its independence of the cosmological background, and discuss a gyroscopic analogue.
Time-dependent stabilization of unbounded potentials based on the Mathieu equation has been discussed in the context of AdS/CFT \cite{Auzzi:2012ca},
a Kapitza-type averaging of an oscillating scale factor in \cite{Kurkov:2016zpd},
and a many-body Kapitza pendulum in \cite{Citro:2015iix}.
Effective potentials induced by an oscillating background have also been discussed for moduli \cite{Lukas:1996wx};
this should be contrasted with the trapping of moduli by particle production \cite{Kofman:2004yc,Green:2009ds}.
For other mechanisms leading to an axion near the hilltop, see \cite{Co:2018mho,Takahashi:2019pqf,Huang:2020etx,Co:2024bme}.

The rest of this paper is organized as follows.
In \cref{sec:Kapitza} we discuss the Kapitza mechanism for an axion with an explicitly time-dependent potential, and analyze it by Floquet theory.
In \cref{sec:dynamical} we discuss models where the oscillation
is generated dynamically, either by a spinning axion or by a massive field oscillating around its minimum.
In \cref{sec:top} we discuss a gyroscopic version of the mechanism
(``inverted spinning top in the sky''), realized by
a nonlinear sigma model on a squashed sphere.
In \cref{sec:backgrounds} we show that the stabilized phase lasts for a number of $e$-folds independent of the cosmological background,
quantify the attractor behavior of the stabilized hilltop,
show that in radiation and matter domination the attractor is ineffective while the Hubble friction is effective but the stabilization can delay the onset of the oscillation,
and discuss the energy budget.
We discuss cosmological applications, including a comment on the swampland conjectures, in \cref{sec:applications}
and conclude in \cref{sec:conclusions}.

%%%%%%%%%%%%%%%%%%%%%%%%%%%%%%%%%%%%%%%%%%%%%%%%%%%%%%%%%%%%
%%%%%%%%%%%%%%%%%%%%%%%%%%%%%%%%%%%%%%%%%%%%%%%%%%%%%%%%%%%%
\section{Kapitza mechanism for axions}\label{sec:Kapitza}
%%%%%%%%%%%%%%%%%%%%%%%%%%%%%%%%%%%%%%%%%%%%%%%%%%%%%%%%%%%%
%%%%%%%%%%%%%%%%%%%%%%%%%%%%%%%%%%%%%%%%%%%%%%%%%%%%%%%%%%%%

%%%%%%%%%%%%%%%%%%%%%%%%%%%%%%%%%%%%%%%%%%%%%%%%%%%%%%%%%%%%
\subsection{Hilltop problem}
%%%%%%%%%%%%%%%%%%%%%%%%%%%%%%%%%%%%%%%%%%%%%%%%%%%%%%%%%%%%

Let us consider a scalar field $\phi$ with potential
\begin{align}
V(\phi) = V_0 (1- \cos \phi).
\label{eq:V_cos}
\end{align}
For example, the (pseudo)scalar $\phi$ is the axion field
(dynamical $\theta$-parameter) for a non-Abelian gauge field \cite{Peccei:1977hh,Weinberg:1977ma,Wilczek:1977pj} (see \cite{Marsh:2015xka} for a review of axion cosmology),
and the potential is generated by non-perturbative instanton effects
breaking the perturbative shift symmetry of $\phi$.
Throughout this paper we normalize the axion by its decay constant $f$, so that
$\phi$ is dimensionless and the Lagrangian density is $f^2 [\frac{1}{2} (\partial \phi)^2 - V(\phi)]$;
in this normalization $V_0=m_{\phi}^2$ has mass dimension two.
The potential is periodic with period $2\pi$, and
has minima at $\phi=2n \pi$ and maxima at $\phi= (2n+1) \pi$ ($n\in \mathbb{Z}$).

In cosmological contexts (to be discussed further in \cref{sec:applications}), we
sometimes wish to consider $\phi$ near the maximum $\phi=\pi$.
In the expanding Universe
with the Hubble parameter $H = \dot{a}/a$,
the homogeneous mode of the scalar field obeys the equation of motion
\begin{align}
\ddot{\phi} + 3 H \dot{\phi} + V'(\phi) = 0.
\label{eq:EOM_phi}
\end{align}
When $H\gg m_{\phi}$, $\phi$ can stay near the maximum of the potential
due to the Hubble friction. Once the Universe cools down
and we have
$H\ll m_{\phi}$, however, a small deviation $\delta\phi = \phi-\pi$
grows as $\delta\phi \propto e^{m_\phi t}$, and
the field quickly rolls down to the bottom of the potential.
Such a hilltop scenario therefore requires a severe fine-tuning of initial conditions.

In the following we discuss a loophole in this argument, which is well known in mechanics as the ``inverted pendulum,'' and adapt it to the cosmological context.

%%%%%%%%%%%%%%%%%%%%%%%%%%%%%%%%%%%%%%%%%%%%%%%%%%%%%%%%%%%%
\subsection{Explicitly oscillating coupling}
%%%%%%%%%%%%%%%%%%%%%%%%%%%%%%%%%%%%%%%%%%%%%%%%%%%%%%%%%%%%

We consider a scalar field $\phi$ with an oscillating potential
\begin{equation}
\label{eq:forced_vibration}
V(\phi,t) = V_0(1- \cos \phi) + b \Omega \sin (\Omega t)\cos \phi ,
\end{equation}
where we assume $\Omega \gg m_\phi, H$.
The second term is the analogue of the vibrating pivot of the Kapitza pendulum:
it modulates the coefficient of $\cos\phi$, i.e.\ the effective ``gravitational acceleration.''

\begin{figure}[t]
\centering
		\includegraphics[width=\columnwidth]{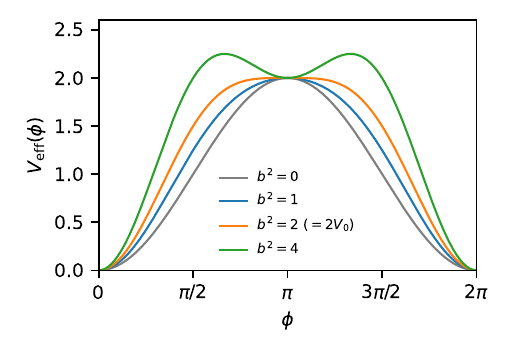}
	\caption{The effective potential $V_{\text{eff}}(\phi) = V_0(1- \cos \phi) + \frac{b^2 }{4} \sin^2 \phi$ for $V_0 = 1$ and representative values of $b$. The hilltop $\phi=\pi$ becomes a local minimum for $b^2>2V_0$.}
	\label{Poteff}
\end{figure}

Since $\Omega$ is large, we can separate $\phi$ into the slow mode
$\phi_s$ and the fast mode $\phi_f$ as
\begin{equation}
\phi(t) = \phi_s(t) + \phi_f(t),
\end{equation}
where $\phi_s(t) = O(1)$ and $\dot{\phi}_s(t) = O(1)$ whereas $\phi_f(t) = O(1/\Omega)$ and $\dot{\phi}_f(t) = O(1)$.
At leading order in $1/\Omega$, the fast mode is driven by the oscillating force
$\ddot{\phi}_f = b\Omega \sin(\Omega t) \sin\phi_s$, and hence
\begin{align}
\phi_f(t) = -\frac{b}{\Omega} \sin (\Omega t) \sin \phi_s .
\label{eq:phi_f}
\end{align}
Substituting this back into \eqref{eq:EOM_phi}, expanding to linear order in $\phi_f$ and averaging over the period $2\pi/\Omega$, we obtain the equation of motion for $\phi_s$:
\begin{align}
\ddot{\phi}_s + 3H \dot{\phi}_s + V_{\text{eff}}'(\phi_s) = 0 ,
\end{align}
with the effective potential
\begin{equation}
V_{\text{eff}}(\phi) = V_0(1- \cos \phi) + \frac{b^2 }{4} \sin^2 \phi ,
\label{eq:Veff}
\end{equation}
see \cref{Poteff}.
This is the standard result that the effective potential is shifted by
$\langle \mathcal{F}^2 \rangle /(2\Omega^2)$, where $\mathcal{F}= b\Omega\sin(\Omega t)\sin\phi$ is the oscillating force \cite{Landau:Mechanics}.
The curvature of the effective potential at $\phi=\pi$ is
\begin{align}
V_{\text{eff}}''(\pi) = \frac{b^2}{2} - V_0 ,
\end{align}
and hence $\phi=\pi$ becomes a metastable local minimum for
\begin{align}
b^2 > 2 V_0 = 2 m_\phi^2 .
\label{eq:stability_Kapitza}
\end{align}
This shows that the addition of the oscillating term in \eqref{eq:forced_vibration}
effectively changes the potential, and hence the positions of its minima.
The slow mode then oscillates around $\phi=\pi$ with frequency $\sqrt{b^2/2-V_0}$, and its amplitude
decays due to the Hubble friction.
Linear stability does not by itself imply capture from an arbitrary initial condition.
For $b^2>2V_0$, the basin of the metastable minimum at $\phi=\pi$ in the averaged potential is bounded by
\begin{align}
\phi_-<\phi<2\pi-\phi_- , \qquad
\phi_-\equiv\arccos\left(-\frac{2V_0}{b^2}\right) .
\label{eq:capture_basin}
\end{align}
Its half-width is $\Delta\phi_{\rm basin}=\arccos(2V_0/b^2)$: it vanishes at the stability threshold and approaches $\pi/2$ for $b^2/V_0\to\infty$.
Nonzero initial velocity and the phase of the fast drive further affect capture.
Thus the mechanism can damp a field already inside a finite basin around the hilltop, but it does not send a completely generic initial condition to $\phi=\pi$.
Note that the consistency of the approximation requires $|\phi_f| \sim b/\Omega \ll 1$,
which together with \eqref{eq:stability_Kapitza} requires
$\Omega \gg b > \sqrt{2} m_{\phi}$.

%%%%%%%%%%%%%%%%%%%%%%%%%%%%%%%%%%%%%%%%%%%%%%%%%%%%%%%%%%%%
\subsection{Floquet analysis} \label{sec:Floquet}
%%%%%%%%%%%%%%%%%%%%%%%%%%%%%%%%%%%%%%%%%%%%%%%%%%%%%%%%%%%%

The effective-potential argument above relies on the separation of time scales.
A more systematic treatment, which does not require $\Omega \gg b$,
is provided by Floquet theory \cite{Floquet:1883}.
Let us linearize around the hilltop, $\phi = \pi + \delta\phi$,
and include also the spatial dependence, $\delta\phi \propto e^{i \vec{k}\cdot \vec{x}}$.
We obtain
\begin{align}
\ddot{\delta\phi}_k + 3H \dot{\delta\phi}_k + \left( \frac{k^2}{a^2} - V_0 + b\Omega \sin(\Omega t)\right)\delta\phi_k = 0 .
\label{eq:linearized}
\end{align}
Let us first ignore the expansion of the Universe and set $a=1$.
By the change of variables $\Omega t = 2\tau - \pi/2$,
\eqref{eq:linearized} becomes the Mathieu equation
\begin{align}
\frac{d^2 \delta\phi_k}{d\tau^2} + \left( \mathsf{a}_k - 2 \mathsf{q} \cos 2\tau \right) \delta\phi_k = 0 ,
\label{eq:Mathieu}
\end{align}
with
\begin{align}
\mathsf{a}_k = \frac{4(k^2 - V_0)}{\Omega^2} , \qquad \mathsf{q} = \frac{2b}{\Omega} .
\end{align}
By Floquet's theorem, the solutions take the form
$\delta\phi_k = e^{\mu_k \tau} P(\tau)$ with $P(\tau)$ periodic with period $\pi$,
and the stability is determined by the Floquet exponent $\mu_k$.
The homogeneous mode is stable if $\mathrm{Re}\,\mu_0 = 0$; as we discuss below, some inhomogeneous modes always lie in instability bands, and their amplification has to be bounded separately.

For the homogeneous mode $k=0$ we have $\mathsf{a}_0<0$: without the oscillation ($\mathsf{q}=0$) this is the familiar tachyonic instability.
The Mathieu equation, however, has a stable region that extends to negative $\mathsf{a}$,
whose boundary is $\mathsf{a} = -\mathsf{q}^2/2 + 7\mathsf{q}^4/128 + O(\mathsf{q}^6)$.
At leading order in $\mathsf{q}$, the stability condition $\mathsf{a}_0 > -\mathsf{q}^2/2$ is precisely the condition \eqref{eq:stability_Kapitza} found from the effective potential.
The Floquet analysis is, however, valid even for $\mathsf{q} = O(1)$.
We have checked numerically that the largest Floquet multiplier of \eqref{eq:linearized} for $k=0$
crosses unity at $b \simeq \sqrt{2 V_0}$ for $\Omega \gg \sqrt{V_0}$.

The Hubble friction can be removed by $\delta\phi_k = a^{-3/2}\, y_k$, after which (for constant $H$)
$y_k$ obeys \eqref{eq:linearized} without the friction term and with $V_0$ replaced by $V_0 + 9H^2/4$.
In the region $b^2>2V_0$, the averaged equation for $y_0$ has the effective mass squared $b^2/2-V_0-9H^2/4>-9H^2/4$, so that $y_0$ grows at most at a rate smaller than $3H/2$, and hence
$\delta\phi_k$ is stable; the condition \eqref{eq:stability_Kapitza} is therefore not modified by $H$.

\begin{figure}[tb]
\centering
		\includegraphics[width=\columnwidth]{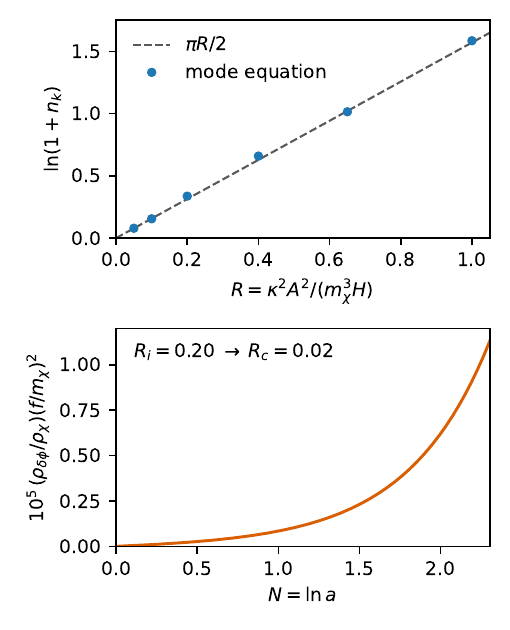}
	\caption{Production and backreaction of inhomogeneous modes.
	Top: direct mode-function evolution across the first resonance band for $b=2$, $\Omega=200$ and $V_0=1$, compared with \eqref{eq:resonance} (dashed line).
	Bottom: the first-band energy-transfer estimate obtained by integrating \eqref{eq:backreaction_fraction} for the delayed-release benchmark $H_i=\tilde m$, $H_c=\tilde m/100$, $m_\chi=10^4\tilde m$ and $r=1/2$, with the factor $(m_\chi/f)^2$ scaled out. This panel uses the local production formula rather than direct evolution of the full spectrum.}
	\label{fig:InhomogeneousBackreaction}
\end{figure}

Sufficiently high-momentum inhomogeneous modes have $\mathsf{a}_k>0$ and can enter the instability bands of the Mathieu equation,
centered at $\mathsf{a}_k \simeq n^2$ $(n=1,2,\dots)$.
This is the familiar parametric resonance \cite{Traschen:1990sw,Kofman:1994rk,Shtanov:1994ce,Kofman:1997yn}.
The first band is located at $k/a \simeq \Omega/2$ with width $\Delta k/a \simeq b$,
and the maximal growth rate there is $\mu_{\rm max}\simeq b/2$
in units of the cosmic time
(we confirmed $\mu_{\rm max} = b/2$ numerically).
In the expanding Universe, a given mode stays in the band only for a time
$\Delta t \simeq 2b/(\Omega H)$.
Near the center of the band the growth rate is $\mu_k \simeq \sqrt{(b/2)^2 - (\omega_k-\Omega/2)^2}$ with $\omega_k \simeq k/a$,
and integrating it along $\dot\omega_k \simeq -H\Omega/2$ gives
$\int \mu_k\, dt \simeq \pi b^2/(4\Omega H)$.
The occupation number of a mode that starts in the vacuum is then amplified to
\begin{align}
n_k \simeq \exp\left[ \frac{\pi}{2} \frac{b^2}{\Omega H} \right] - 1 ,
\qquad
|\delta\phi_k| \propto \exp\left[ \frac{\pi}{4} \frac{b^2}{\Omega H} \right] .
\label{eq:resonance}
\end{align}
We have confirmed \eqref{eq:resonance} by numerically solving \eqref{eq:linearized}
across the first band with $a = e^{Ht}$: for $b^2/(\Omega H) = 3.2,\ 6.4,\ 12.8$ we find $\ln(1+n_k) = 5.1,\ 10.1,\ 20.3$,
compared with $5.0,\ 10.1,\ 20.1$ from \eqref{eq:resonance}, and the result depends on $b,\Omega,H$ only through $b^2/(\Omega H)$.
The upper panel of \cref{fig:InhomogeneousBackreaction} provides a complementary check in the weak-production regime relevant for avoiding backreaction: direct mode-function evolution agrees with $\ln(1+n_k)=\pi b^2/(2\Omega H)$ within $8\%$ for $0.05\le b^2/(\Omega H)\le1$.
For $b^2/(\Omega H)\lesssim 1$ the band is crossed non-adiabatically and $n_k = \mathcal{O}(1)$.
The resonance is therefore inefficient when $b^2 \lesssim \Omega H$;
combined with \eqref{eq:stability_Kapitza}, the hilltop can be stabilized without exciting the inhomogeneous modes if
$2 m_\phi^2 < b^2 \lesssim \Omega H$.
This criterion ($n_k\lesssim 1$) is sufficient but not necessary: what is actually required is that the energy density and variance of the amplified modes remain too small to backreact on the homogeneous mode.
The corresponding upper bound on $b^2/(\Omega H)$ depends logarithmically on the initial fluctuation spectrum, the duration of the drive and the available energy density, and is therefore model dependent.
The resonant momentum $k/a\simeq\Omega/2$ must also lie below the cutoff of the effective theory; conversely, using $\Omega$ above the cutoff would invalidate the periodically driven description rather than provide a way to remove the resonance.
The numerical solutions shown below follow only the homogeneous mode.

%%%%%%%%%%%%%%%%%%%%%%%%%%%%%%%%%%%%%%%%%%%%%%%%%%%%%%%%%%%%
\section{Dynamical realizations}\label{sec:dynamical}
%%%%%%%%%%%%%%%%%%%%%%%%%%%%%%%%%%%%%%%%%%%%%%%%%%%%%%%%%%%%

In the previous section the oscillating term of the potential was introduced by
hand. Here we consider models where the oscillation originates from
the dynamics of another scalar field.

%%%%%%%%%%%%%%%%%%%%%%%%%%%%%%%%%%%%%%%%%%%%%%%%%%%%%%%%%%%%
\subsection{Spinning axion}\label{sec:spinning}
%%%%%%%%%%%%%%%%%%%%%%%%%%%%%%%%%%%%%%%%%%%%%%%%%%%%%%%%%%%%

We consider two axion fields $\phi, \chi$ with the potential
\begin{equation}
V(\phi,\chi) = V_0(1- \cos \phi) - b \omega\cos(\chi/g) \cos \phi .  \label{eq:CoupledAxions}
\end{equation}
Here we introduced a constant $g\gg 1$ so that the back-reaction on $\chi$, whose force is $\propto 1/g$, is small.
We consider a fast-spinning initial condition for the $\chi$ field.
More explicitly, a sufficient condition for the interaction not to appreciably change the rotation during one period is that the rotational kinetic energy dominate the modulation,
$g^2\Omega^2\gg b\omega$.
This condition is satisfied throughout the stability window below when $g\gg1$, but it should be imposed independently in a general parameter scan.
Ignoring the interaction with $\phi$, the equation of motion $\ddot\chi + 3H\dot\chi=0$ gives (for constant $H$)
\begin{equation}
\dot\chi(t) = \dot\chi_0\, e^{-3H t} , \qquad
\chi(t) =  \chi_0 + \frac{\dot\chi_0}{3H}\left(1- e^{-3H t}\right) .
\end{equation}
The coupling in \eqref{eq:CoupledAxions} then oscillates with the instantaneous angular frequency
\begin{align}
\Omega(t) \equiv \frac{\dot\chi(t)}{g} = \Omega_0\, e^{-3Ht} ,
\end{align}
which is slowly varying as long as $H \ll \Omega$.
Comparing with \eqref{eq:forced_vibration}, the amplitude $b\Omega$ is now replaced by $b\omega$,
and the slow mode of $\phi$ feels the effective potential
\begin{align}
V_{\text{eff}}(\phi) &= V_0(1- \cos \phi) + \frac{b^2  \omega^2}{4\,\Omega(t)^2} \sin^2 \phi \notag\\
&= V_0(1- \cos \phi) + \frac{b^2  \omega^2 g^2}{4\,\dot{\chi}^2} \sin^2 \phi .
\end{align}
The stability condition \eqref{eq:stability_Kapitza} now reads $\Omega(t) < \Omega_+ \equiv b\omega/\sqrt{2V_0}$.
Interestingly, the stabilizing term \emph{grows} as $\chi$ slows down.
The slow-down, however, eventually invalidates the approximation:
the Mathieu parameter $\mathsf{q} = 2b\omega/\Omega^2$ grows,
and $\delta\phi$ crosses the boundary of the first stability region of the Mathieu equation at $\mathsf{q}\simeq 0.91$ (for $|\mathsf{a}|\ll 1$) \cite{DLMF},
i.e.\ when $\Omega(t)$ reaches $\Omega_- \simeq (2 b\omega/0.91)^{1/2}$, after which $\phi$ leaves the hilltop.
The stabilized phase thus exists only if $\Omega_+ > \Omega_-$, i.e.\ $b\omega \gtrsim 4.4\,V_0$, and
lasts for
\begin{align}
\Delta N \simeq \frac{1}{3} \ln \frac{\Omega_+}{\Omega_-} \simeq \frac{1}{6} \ln\frac{0.91\, b\omega}{4V_0}
\label{eq:DeltaN}
\end{align}
$e$-folds. This is logarithmic in the hierarchy $b\omega/V_0$,
reflecting the fact that the kinetic energy of the spinning axion redshifts as $a^{-6}$.
Note also that the resonance parameter of \cref{sec:Floquet}, $b_{\rm eff}^2/(\Omega H)$ with $b_{\rm eff}=b\omega/\Omega$, grows as $\Omega$ decreases, so that the condition for the absence of parametric resonance becomes more stringent towards the end of the window.
Before the stabilized phase, i.e.\ for $\Omega(t)>\Omega_+$, the hilltop is unstable but $\phi$ can be kept near the top by the Hubble friction if $H \gtrsim m_\phi$.

\begin{figure}[t]
\centering
		\includegraphics[width=\columnwidth]{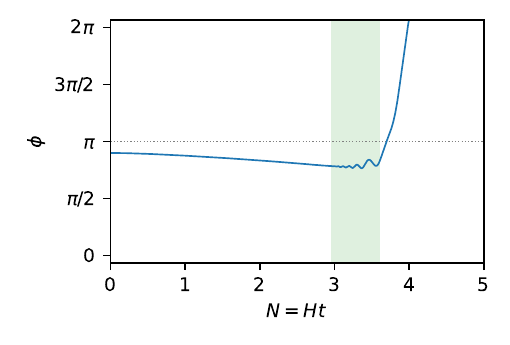}
	\caption{Time evolution of $\phi$ for the potential \eqref{eq:CoupledAxions},
 with $V_0 = 1, b=2, \omega = 100, g = 100$ and $H=1$.
 The initial conditions are  $\phi = \frac{9}{10}\pi, \chi= 0$ and $\dot{\phi} = 0, \dot{\chi} = 10^6 g$ ($\Omega_0=10^6$).
 The shaded region is the stabilized phase $\Omega_-<\Omega(t)<\Omega_+$ predicted in the text.}
	\label{TwoAxionTE}
\end{figure}

\begin{figure}[t]
\centering
		\includegraphics[width=\columnwidth]{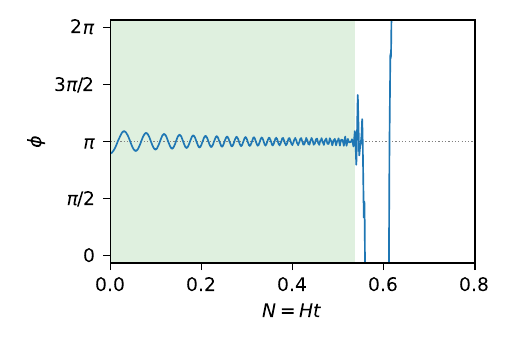}
	\caption{Same as \cref{TwoAxionTE}, but for $H=0.01$ and $\dot\chi(0) = 10^2 g$ ($\Omega_0=100$).}
	\label{TwoAxionSH}
\end{figure}

We solve the time evolution under the potential \eqref{eq:CoupledAxions} numerically and show the results in \cref{TwoAxionTE,TwoAxionSH}, for $V_0 = 1$, $b=2$, $\omega = 100$, $g = 100$, so that $\Omega_+ \simeq 141$ and $\Omega_- \simeq 21$.
In \cref{TwoAxionTE} we take $H=1$ and $\Omega_0 = 10^6 \gg \Omega_+$.
The hilltop is initially unstable, but $\phi$ is held by the Hubble friction (only marginally, since $H=m_\phi$, so that $\phi$ slowly drifts from its initial value);
the stabilized phase starts at $N\simeq \frac{1}{3}\ln(\Omega_0/\Omega_+)\simeq 3.0$ and ends at $N \simeq \frac{1}{3}\ln(\Omega_0/\Omega_-) \simeq 3.6$,
and $\phi$ indeed leaves the hilltop at $N\simeq 3.8$.
In \cref{TwoAxionSH} we take $H=0.01$ and $\Omega_0 = 100$, so that the hilltop is stabilized from the beginning.
The field $\phi$ oscillates around $\phi=\pi$, with an amplitude decreasing due to the Hubble friction and a frequency $\sqrt{b^2\omega^2/(2\Omega^2)-V_0}$ increasing as $\Omega$ decreases.
The stabilization is lost at $N \simeq \frac{1}{3}\ln(\Omega_0/\Omega_-) \simeq 0.52$, and $\phi$ leaves the hilltop at $N\simeq 0.54$, in agreement with \eqref{eq:DeltaN}.

These figures show the homogeneous mode only.
For the spinning axion, the parametric resonance of inhomogeneous modes cannot be avoided towards the end of the window:
at $\Omega=\Omega_-$ the resonance parameter is $b_{\rm eff}^2/(\Omega_- H)=b^2\omega^2/(\Omega_-^3H)\simeq (b\omega)^{1/2}/(3.3\,H)$,
so that the sufficient condition $b_{\rm eff}^2\lesssim\Omega H$ requires $H\gtrsim 0.3\,(b\omega)^{1/2}$, which exceeds $0.6\,m_\phi$ for $b\omega\gtrsim 4.4\,V_0$;
in this regime the Hubble friction alone would hold the field near the hilltop.
For the parameters of \cref{TwoAxionTE,TwoAxionSH}, the resonance parameter at $\Omega_-$ is about $4$ and $4\times 10^2$, respectively.
Moreover, since the homogeneous mode leaves the stability region at $\Omega_-$ through the edge of the first instability band, modes with $k\neq 0$, which have a larger $\mathsf{a}$, enter this band even before $\Omega$ reaches $\Omega_-$.
The end of the stabilized phase of the spinning axion is therefore expected to be accompanied by the growth of inhomogeneous modes, and to be even more violent than the ejection of the homogeneous mode seen in the figures.

%%%%%%%%%%%%%%%%%%%%%%%%%%%%%%%%%%%%%%%%%%%%%%%%%%%%%%%%%%%%
\subsection{Oscillating massive field}\label{sec:clean}
%%%%%%%%%%%%%%%%%%%%%%%%%%%%%%%%%%%%%%%%%%%%%%%%%%%%%%%%%%%%

\begin{figure}[t]
\centering
		\includegraphics[width=\columnwidth]{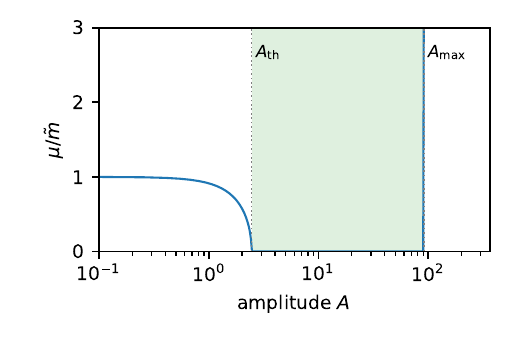}
	\caption{The Floquet exponent $\mu$ (growth rate per unit cosmic time) of $\delta\phi$ around $\phi=\pi$ for the model \eqref{eq:clean} with $V_0=1$, $\kappa=50$, $m_\chi=100$ and $H=0$, as a function of the amplitude $A$ of $\chi$.
	The hilltop is linearly stable ($\mu=0$) in the shaded region $A_{\rm th}<A<A_{\rm max}$. For $A\to 0$, $\mu$ approaches the tachyonic rate $\tilde m$; for $A>A_{\rm max}$ the mode enters the first instability band of the Mathieu equation and $\mu$ grows rapidly.}
	\label{fig:FloquetA}
\end{figure}

Instead of a spinning $\chi$, we can also consider $\chi$ oscillating around the minimum of its own potential.
The simplest model in which the stabilization is purely dynamical is
\begin{align}
V(\phi,\chi) = V_0(1-\cos\phi) - \kappa\, \chi \cos\phi + \frac{1}{2} m_\chi^2 \chi^2 .
\label{eq:clean}
\end{align}
For $m_\chi \gg H$, $\chi$ oscillates as $\chi = \bar\chi + A \cos(m_\chi t)$ with $A\propto a^{-3/2}$,
where $\bar\chi = -\kappa/m_\chi^2$ at $\phi=\pi$.
The fluctuation $\delta\phi$ around the hilltop then obeys the Mathieu equation \eqref{eq:Mathieu}
with $\mathsf{a}_0 = -4\tilde{m}^2/m_\chi^2$ and $\mathsf{q}=2\kappa A/m_\chi^2$, where
$\tilde m^2 \equiv m_\phi^2 - \kappa^2/m_\chi^2$.
We assume $\kappa < m_\chi m_\phi$, so that $\tilde m^2>0$: the hilltop is a maximum of the potential averaged over the oscillation (although the instantaneous curvature $-(\tilde m^2+\kappa A\cos m_\chi t)$ changes sign for $\kappa A>\tilde m^2$, as for the Kapitza pendulum whose pivot acceleration exceeds the gravitational acceleration),
and the stabilization is entirely due to the oscillation.
The hilltop is stable for
\begin{align}
A_{\rm th} < A < A_{\rm max} , \notag\\
A_{\rm th} = \frac{\sqrt{2}\, m_\chi \tilde{m}}{\kappa} , \quad
A_{\rm max} \simeq \frac{0.91\, m_\chi^2}{2\kappa} ,
\end{align}
where the lower bound is the Kapitza condition \eqref{eq:stability_Kapitza} and the upper bound
is the edge $\mathsf{q}\simeq 0.91$ of the first stability region of the Mathieu equation,
see \cref{fig:FloquetA}, where we show the Floquet exponent computed numerically.
The stabilized phase thus lasts for
\begin{align}
\Delta N = \frac{2}{3}\ln\frac{A_i}{A_{\rm th}} = \frac{1}{3}\ln\frac{\rho_{\chi,i}}{\rho_{\rm th}}
\le \frac{2}{3} \ln\left( 0.32\, \frac{m_\chi}{\tilde{m}}\right) ,
\label{eq:DeltaN_clean}
\end{align}
where $\rho_\chi\propto A^2$ is the energy density of the oscillation of $\chi$.
In terms of dimensionful quantities, \eqref{eq:clean} corresponds to the Lagrangian
\begin{align}
\mathcal{L} &= \frac{f^2}{2}(\partial\phi)^2 + \frac{1}{2}(\partial X)^2 - f^2 m_\phi^2 (1-\cos\phi) \notag\\
&\quad + \Lambda_X^3 X\cos\phi - \frac{1}{2}m_\chi^2 X^2 ,
\label{eq:clean_dim}
\end{align}
with a canonically normalized $X = f\chi$, so that $V_0=m_\phi^2$, $\kappa = \Lambda_X^3/f$, and $f A$ is the physical amplitude of $X$.
Such a coupling arises, e.g., when the coefficient of the instanton-induced potential of $\phi$ depends on another field, $\Lambda^4(X)\cos\phi$ with $\Lambda^4(X)\simeq\Lambda_0^4+\Lambda_X^3 X$.
The condition $\kappa<m_\chi m_\phi$ reads $\Lambda_X^3< f m_\chi m_\phi$, and the energy density of the oscillation at the threshold is
\begin{align}
\rho_{\rm th} = \frac{1}{2}f^2 m_\chi^2 A_{\rm th}^2 = \frac{1-r^2}{r^2}\, f^2 m_\chi^2 , \qquad r\equiv\frac{\kappa}{m_\chi m_\phi}<1 .
\label{eq:rho_th}
\end{align}
Since $A\propto a^{-3/2}$ for any $H(t)\ll m_\chi$, the stabilized phase lasts for a fixed number of $e$-folds, independently of the cosmological background; we come back to this point in \cref{sec:backgrounds}.

\paragraph{Regime of validity.}
The two-field effective description requires, in addition to the Floquet window $A_{\rm th}<A<A_{\rm max}$,
\begin{align}
H,\tilde m &\ll m_\chi, &
\rho_\chi &\lesssim 3M_P^2H^2, &
\frac{\kappa^2A^2}{m_\chi^3H}&\lesssim 1
\label{eq:validity}
\end{align}
if one uses the sufficient condition that the first resonance band be inefficient.
The physical amplitude $fA$ must remain within the range in which the expansion $\Lambda^4(X)\simeq\Lambda_0^4+\Lambda_X^3X$ is valid, and $m_\chi$ and the resonant momentum $m_\chi/2$ must lie below the cutoff of the effective theory.
These requirements, together with the available energy budget, can be more restrictive than linear stability of the homogeneous mode.

Let us comment on the importance of the condition $\kappa < m_\chi m_\phi$.
If it is violated, the static shift $\bar\chi$ alone makes $\phi=\pi$ a minimum of the potential, and no oscillation is needed.
More generally, for a coupling $-B(\chi)\cos\phi$ the linearized equation around $\phi=\pi$ is the Hill equation
\begin{align}
\ddot{\delta\phi} + 3H\dot{\delta\phi} - \left( V_0 + B(t) \right)\delta\phi = 0 ,
\label{eq:Hill}
\end{align}
and writing $B(t) = \bar{B} + B_1 \cos(\Omega t) + \dots$, the leading-order stability condition is $B_1^2/(2\Omega^2) > V_0 + \bar{B}$.
Both models of this section are of this form:
for the spinning axion \eqref{eq:CoupledAxions}, $B(t)=b\omega\cos(\chi(t)/g)$ with $\chi\simeq g\Omega t$, so that $\bar B\simeq 0$ and $B_1=b\omega$;
for \eqref{eq:clean}, $B(t)=\kappa\chi(t)$, so that $\bar B=-\kappa^2/m_\chi^2$ and $B_1=\kappa A$ with $\Omega=m_\chi$.
The stabilization is genuinely dynamical only if $V_0+\bar B>0$.
For instance, for a periodic coupling $-b\omega\cos(\chi/g)\cos\phi$ with $\chi$ oscillating around a minimum of its own potential where $\cos(\chi/g)<0$,
we find numerically, in a non-hierarchical example ($V_0=1$, $b=\omega=g=2$ and a potential $m_\chi^2(1-\cos\chi)$ with $m_\chi^2=3.4$, at $H=0$), that the hilltop is linearly stable only when $V_0+B(t)<0$ throughout the oscillation, i.e.\ when the hilltop is a minimum at any instant.
This reflects the absence of a frequency hierarchy in this example: a genuinely dynamical stabilization requires $\Omega\gg\tilde m$, as in \eqref{eq:clean} with $m_\chi\gg m_\phi$.
The sign flip of the averaged potential discussed in \cite{Murai:2025wbg}, where $\bar B$ is the average of a modulated cosine (a Bessel function of the oscillation amplitude), corresponds to the case $V_0+\bar B<0$.

%%%%%%%%%%%%%%%%%%%%%%%%%%%%%%%%%%%%%%%%%%%%%%%%%%%%%%%%%%%%
\section{Inverted spinning top in the sky}\label{sec:top}
%%%%%%%%%%%%%%%%%%%%%%%%%%%%%%%%%%%%%%%%%%%%%%%%%%%%%%%%%%%%

There is another classic way to stabilize an inverted mechanical system: a fast-spinning top can stand upright (the ``sleeping top'').
In contrast with the Kapitza pendulum, the stabilization here is due to the conserved angular momentum (gyroscopic stabilization) rather than the time-periodic driving.

We consider scalar fields $\Phi^i$ described by the nonlinear sigma model
\begin{equation}
S = \int d^4 x \sqrt{-g} \left[ -\frac{1}{2} g^{\mu\nu} G_{ij}(\Phi) \partial_{\mu} \Phi^i \partial_{\nu} \Phi^j - V(\Phi) \right] ,
\end{equation}
where we use the mostly-plus signature.
To consider a motion similar to the spinning top, we take the target space to be the squashed sphere $\tilde{S}^3$
with Euler angles $\Phi^i = (\vartheta,\varphi,\psi)$
and the metric
\begin{align}
ds_{\Phi}^2 &= G_{ij} d\Phi^i d\Phi^j  \notag \\
&= I_1 (d\vartheta^2 + \sin^2 \vartheta\, d\varphi^2 ) + I_3 (d\psi+ \cos \vartheta\, d\varphi)^2 ,
\end{align}
where the constants $I_1, I_3$ have mass dimension two and play the roles of the moments of inertia.
For the potential we take the analogue of the gravitational potential of the top,
\begin{align}
V(\vartheta) = \mathcal{V}_0 (1 + \cos\vartheta) ,
\end{align}
where $\mathcal{V}_0$ has mass dimension four and is distinguished from the mass-squared parameter $V_0$ used in the axion sections.
The maximum at $\vartheta=0$ corresponds to the upright top.
The equations of motion are
\begin{equation}
\Box \Phi^i + \Gamma(\Phi)^i_{jk}\, g^{\mu\nu} \partial_{\mu}\Phi^j\partial_{\nu}\Phi^k - G^{ij}\partial_j V  = 0 .
\end{equation}
In a flat FLRW background and focusing on the homogeneous motion, we get
\begin{equation}
\ddot{\Phi}^i + 3 H(t)\dot{\Phi}^i + \Gamma(\Phi)^i_{jk} \dot{\Phi}^j \dot{\Phi}^k + G^{ij}\partial_j V = 0 .
\end{equation}
Since $\varphi$ and $\psi$ are cyclic, it is convenient to use the Lagrangian per comoving volume,
$L = a^3 [\frac{1}{2}G_{ij}\dot\Phi^i\dot\Phi^j - V]$.
The momenta conjugate to $\psi$ and $\varphi$ are conserved,
which means that the physical momentum densities
\begin{align}
p_\psi &\equiv I_3 (\dot\psi + \cos\vartheta\, \dot\varphi) , \\
p_\varphi &\equiv I_1 \sin^2\vartheta\, \dot\varphi + p_\psi\cos\vartheta
\end{align}
both scale as $a^{-3}$.
Eliminating $\dot\varphi, \dot\psi$, the equation of motion for $\vartheta$ becomes
\begin{align}
I_1 (\ddot\vartheta + 3H\dot\vartheta) = - \frac{\partial U}{\partial\vartheta} ,
\end{align}
with the effective potential (Routhian)
\begin{align}
U(\vartheta) = \frac{(p_\varphi - p_\psi \cos\vartheta)^2}{2 I_1 \sin^2\vartheta} + V(\vartheta) .
\end{align}
For the ``sleeping top'' at $\vartheta=0$, regularity requires $p_\varphi = p_\psi$
(a condition preserved by the expansion), and expanding around $\vartheta=0$ we find
\begin{align}
U(\vartheta) = 2\mathcal{V}_0 + \frac{1}{2}\left( \frac{p_\psi^2}{4 I_1} - \mathcal{V}_0 \right)\vartheta^2 + O(\vartheta^4) .
\end{align}
The hilltop is therefore stabilized for
\begin{align}
p_\psi^2 > 4 I_1 \mathcal{V}_0 ,
\label{eq:top_stability}
\end{align}
which is the field-theory analogue of the well-known condition $I_3^2\omega_3^2 > 4 I_1 M g l$ for the sleeping top.
Since $p_\psi \propto a^{-3}$, this condition holds only for
\begin{align}
\Delta N = \frac{1}{3}\ln \frac{p_{\psi}(0)}{2\sqrt{I_1 \mathcal{V}_0}}
\label{eq:DeltaN_top}
\end{align}
$e$-folds, again logarithmic in the hierarchy, as in \eqref{eq:DeltaN}.

\begin{figure}[tb]
\centering
		\includegraphics[width=\columnwidth]{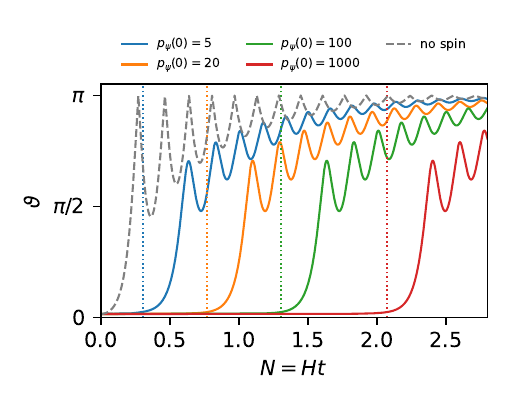}
	\caption{Numerical solutions of the inverted spinning top for $I_1=I_3=\mathcal{V}_0=1$, $H=0.05$, $\vartheta(0)=0.05$ and several values of $p_\psi(0)$.
	The dotted vertical lines show the predicted end of the stabilized phase, \eqref{eq:DeltaN_top}. The dashed curve is the solution without spin.}
	\label{fig:SpinningTop}
\end{figure}

We have solved the full equations of motion numerically, see \cref{fig:SpinningTop}.
Since the squashed $S^3$ with the metric above is the configuration space of a symmetric rigid body,
it is convenient to integrate the equivalent rigid-body equations
for the orientation (parametrized by a unit quaternion) and the angular momentum $\vec{L}$ in the space frame,
$\dot{\vec{L}} = -\mathcal{V}_0\, \vec{n}\times \hat{z} - 3H\vec{L}$, where $\vec n$ is the symmetry axis and $\cos\vartheta = \vec{n}\cdot\hat z$;
this avoids the coordinate singularity of the Euler angles at $\vartheta=0$.
Starting from $\vartheta(0)=0.05$ with $\vec{L}$ parallel to $\vec{n}$ (so that $p_\varphi=p_\psi\cos\vartheta(0)$, i.e.\ $p_\varphi=p_\psi$ up to $\mathcal{O}(\vartheta(0)^2)$),
the field stays at $\vartheta \lesssim 0.08$ until the predicted $e$-fold \eqref{eq:DeltaN_top},
and falls from the hilltop $0.25$ $e$-folds after that for all values of $p_\psi(0)$ we studied, whereas without the spin it falls immediately.
In contrast with the Kapitza case, the deviation $\vartheta$ is not damped towards the hilltop but stays of the order of its initial value (it grows slowly, from $0.05$ to $0.08$, partly because of the small nutation induced by $p_\varphi\neq p_\psi$ at this order), so that the sleeping top holds the field rather than attracting it.
Note also that the stability condition \eqref{eq:top_stability} is independent of $I_3$, so that the numerical results for $I_1=I_3$ shown in \cref{fig:SpinningTop} apply also to the squashed case once $p_\psi$ is fixed.

Let us comment on the origin of the sigma model.
For $I_1=I_3$ the target space is the round $S^3 \simeq SU(2)$, and the model is the $SU(2)$ principal chiral model with the $SU(2)_L\times SU(2)_R$ symmetry.
The squashed metric with $I_1\neq I_3$ is the most general metric invariant under $SU(2)_L\times U(1)_R$,
and hence arises generically once the right $SU(2)$ is broken to $U(1)$, e.g.\ by a vacuum expectation value of a field in the adjoint representation.
The potential $V(\vartheta)$ further breaks the symmetry down to $U(1)\times U(1)$, generated by the shifts of $\varphi$ and $\psi$,
whose conserved charges are $p_\varphi$ and $p_\psi$ above.
The spinning state $p_\psi\neq 0$ is thus a state with a nonzero charge density of a $U(1)$ symmetry, similar to the axion rotations discussed in the context of axiogenesis and the kinetic misalignment mechanism \cite{Co:2019wyp,Co:2019jts}.
In terms of the $SU(2)$-valued field $U$, a potential of the form $V(\vartheta)$ is generated, e.g., by a term $\propto\mathrm{Tr}(\sigma_3 U\sigma_3U^\dagger)=2\cos\vartheta$, analogous to the electromagnetic mass term of pions in the chiral Lagrangian.
In cosmology, angular motion in field space supporting a field against its potential has been discussed in hyperinflation \cite{Brown:2017osf} and spinflation \cite{Easson:2007dh}, and $SU(2)$-valued background fields appear in chromo-natural inflation and gauge-flation \cite{Adshead:2012kp,Maleknejad:2011jw};
we are not aware of a discussion of a field held at a maximum of its potential by angular momentum (the ``sleeping top'') in cosmology.

%%%%%%%%%%%%%%%%%%%%%%%%%%%%%%%%%%%%%%%%%%%%%%%%%%%%%%%%%%%%
\section{Cosmological backgrounds and the hilltop attractor}\label{sec:backgrounds}
%%%%%%%%%%%%%%%%%%%%%%%%%%%%%%%%%%%%%%%%%%%%%%%%%%%%%%%%%%%%

So far we have mostly assumed a constant Hubble parameter.
In this section we discuss how the results depend on the cosmological background,
inflation ($H=$ const.), radiation domination ($a\propto t^{1/2}$) and matter domination ($a\propto t^{2/3}$),
and quantify how efficiently the stabilized hilltop attracts the field.

%%%%%%%%%%%%%%%%%%%%%%%%%%%%%%%%%%%%%%%%%%%%%%%%%%%%%%%%%%%%
\subsection{Universality of the stabilized phase}
%%%%%%%%%%%%%%%%%%%%%%%%%%%%%%%%%%%%%%%%%%%%%%%%%%%%%%%%%%%%

In all the examples above, the agent of the stabilization is diluted by the expansion with a fixed power of the scale factor:
the angular velocity of the spinning axion scales as $\Omega\propto a^{-3}$ (\cref{sec:spinning}),
the amplitude of an oscillating massive field as $A\propto a^{-3/2}$ (\cref{sec:clean}),
and the angular momentum density of the spinning top as $p_\psi\propto a^{-3}$ (\cref{sec:top}).
These scalings hold for any $H(t)$, as long as $H$ is small compared with the frequency of the agent.
Since the stability conditions are fixed thresholds on these quantities,
the stabilized phase begins and ends at fixed values of the scale factor, and its duration
$\Delta N$, as given in \eqref{eq:DeltaN}, \eqref{eq:DeltaN_clean} and \eqref{eq:DeltaN_top},
is \emph{independent of the cosmological background}.
We have confirmed this numerically for the spinning axion \eqref{eq:CoupledAxions} with the parameters of \cref{TwoAxionSH}:
the field leaves the hilltop at $N= 0.543$, $0.548$ and $0.539$ for de Sitter, radiation and matter domination, respectively
(starting from $H=0.01$ in all cases), compared with the end of the stabilized phase $N=\frac13\ln(\Omega_0/\Omega_-)= 0.52$ predicted above.

The background nevertheless matters in two respects.
First, for $a\propto t^p$ a fixed number of $e$-folds corresponds to a much longer physical time,
$t_{\rm end}/t_{\rm start} = e^{\Delta N/p}$, i.e.\ a power of the hierarchy:
for instance \eqref{eq:DeltaN} gives $t_{\rm end}/t_{\rm start} \simeq (0.91\,b\omega/4V_0)^{1/(6p)}$,
which is $(0.91\,b\omega/4V_0)^{1/3}$ for radiation domination.
Second, the fate of the field outside the stabilized phase depends on the ratio $H/m_\phi$.
During inflation with $H\gtrsim m_\phi$ the field is held by the Hubble friction both before and after the stabilized phase,
while in radiation and matter domination $H$ decreases and the field rolls down within a Hubble time once it is released at $H\ll m_\phi$
(see \cref{fig:Backgrounds}).
In the decelerating Universe, therefore, the stabilized phase should start before $H$ drops below $m_\phi$, i.e.\
within $N\lesssim p\ln(H_{i}/m_\phi)$ $e$-folds after the initial time with Hubble parameter $H_i$.

%%%%%%%%%%%%%%%%%%%%%%%%%%%%%%%%%%%%%%%%%%%%%%%%%%%%%%%%%%%%
\subsection{The hilltop attractor}\label{sec:attractor}
%%%%%%%%%%%%%%%%%%%%%%%%%%%%%%%%%%%%%%%%%%%%%%%%%%%%%%%%%%%%

\begin{figure}[t]
\centering
		\includegraphics[width=\columnwidth]{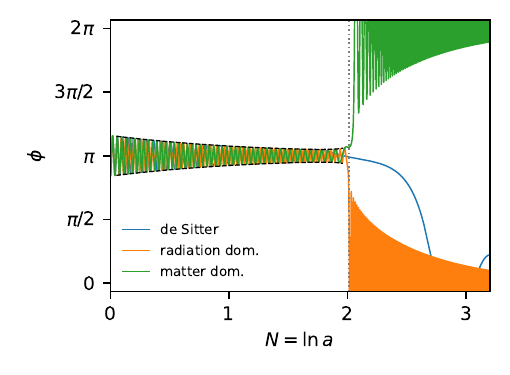}
	\caption{Time evolution of $\phi$ for the model \eqref{eq:clean} with $V_0=1$, $\kappa=5\times 10^3$, $m_\chi=10^4$,
	in de Sitter, radiation and matter domination, all starting from $H=0.1$ (so that $H<\tilde m$ throughout, and the condition \eqref{eq:noresonance} is satisfied),
	with $\phi(0)=\pi-0.5$, $\chi(0)=\bar\chi+50$ and zero initial velocities.
	The dashed curves show the adiabatic envelope $|\delta\phi|\propto a^{-3/2}\omega_{\rm eff}^{-1/2}$,
	and the dotted line the predicted end of the stabilized phase \eqref{eq:DeltaN_clean}.}
	\label{fig:Backgrounds}
\end{figure}

To make quantitative statements, we now consider the model \eqref{eq:clean} of \cref{sec:clean}, in which the stabilization is purely dynamical.
During the stabilized phase, the hilltop is an \emph{attractor}:
$\delta\phi$ oscillates with frequency $\omega_{\rm eff}^2 = \kappa^2 A^2/(2m_\chi^2) - \tilde m^2$,
and its amplitude decreases as $|\delta\phi| \propto a^{-3/2}\omega_{\rm eff}^{-1/2}$ (adiabatic invariant).
As shown in \cref{fig:Backgrounds}, this agrees well with the numerical solutions, again independently of the background.
The adiabatic damping, however, operates only as long as $\omega_{\rm eff}\gtrsim H$, and the final deviation depends on whether the stabilized phase ends after or before the Hubble friction becomes ineffective.

\paragraph{Release at $H_c\ll\tilde m$.}
In this case $\omega_{\rm eff}$ goes to zero at the end of the stabilized phase, and the adiabatic approximation breaks down.
This transition can be treated analytically.
Since $\omega_{\rm eff}^2 = \tilde m^2 (A^2/A_{\rm th}^2-1) = \tilde m^2(e^{3(N_c-N)}-1)$, where $N_c$ is the end of the stabilized phase,
near $t=t_c$ we have $\omega_{\rm eff}^2 \simeq 3\tilde m^2 H_c (t_c-t)$, with $H_c$ the Hubble parameter at $t_c$.
For $x\equiv a^{3/2}\delta\phi$ (which removes the Hubble friction up to corrections of order $H^2$), the slow mode then obeys the Airy equation
\begin{align}
\frac{d^2x}{ds^2} = s\, x , \qquad s \equiv \frac{t-t_c}{\tau_c} , \qquad \tau_c \equiv (3\tilde m^2 H_c)^{-1/3} ,
\end{align}
which is valid for $H_c\tau_c \ll 1$, i.e.\ $H_c\ll\tilde m$.
Matching to the adiabatic solution $x = Y\cos\Theta$ with $Y\propto\omega_{\rm eff}^{-1/2}$ for $s\ll -1$,
we obtain $x = \sqrt{\pi}\,\mathcal{A}\left[\sin\Theta_c\, \mathrm{Ai}(s) + \cos\Theta_c\, \mathrm{Bi}(s)\right]$,
with $\mathcal{A} = Y_i(\omega_{\rm eff,i}\tau_c)^{1/2}$ and $\Theta_c$ the phase of the oscillation at the transition.
For $s\gg 1$ this is the tachyonic growing mode, $x\simeq \mathcal{A}\cos\Theta_c\, s^{-1/4}e^{\frac{2}{3}s^{3/2}}$.
We therefore define the residual deviation at the release by the amplitude of this mode, $\delta_{\rm rel}\equiv \mathcal{A}/a_c^{3/2}$:
\begin{align}
\delta_{\rm rel} &= \mathcal{C}\, \delta_i\, e^{-3\Delta N/4} = \mathcal{C}\, \delta_i \left(\frac{\rho_{\rm th}}{\rho_{\chi,i}}\right)^{1/4} , \notag\\
\mathcal{C} &= (\tilde m \tau_c)^{1/2}\left(1-e^{-3\Delta N}\right)^{1/4} \simeq \left(\frac{\tilde m}{3H_c}\right)^{1/6} ,
\label{eq:delta_rel}
\end{align}
where $\delta_i$ is the initial deviation.
The deviation that actually seeds the roll-down is $\delta_{\rm rel}|\cos\Theta_c|$;
the phase $\Theta_c\simeq\int\omega_{\rm eff}\,dt$ is a rapidly varying function of the parameters, and $\delta_{\rm rel}$ is its envelope.
We have checked \eqref{eq:delta_rel} by solving the full equations of motion for $\phi$ and the oscillating $\chi$ (without the averaging) with a tiny initial deviation, in radiation and matter domination, for $\Delta N = 0.9$--$2.8$ (for which $\mathcal{C} = 1.5$--$3.0$):
the envelope of $x$ over the initial phases agrees with $\sqrt{\pi}\mathcal{A}\,[\mathrm{Ai}(s)^2+\mathrm{Bi}(s)^2]^{1/2}$ within $10\%$ for $|s|\le 2$.
In this comparison it is important to prepare the initial state of the slow mode including the fast mode \eqref{eq:phi_f}, whose velocity is of the same order as that of the slow mode.
Combined with $A_i<A_{\rm max}$, the precision attainable in this case is
\begin{align}
\delta_{\rm rel} \gtrsim \mathcal{C}\,\delta_i \left( \frac{3\, \tilde{m}}{m_\chi}\right)^{1/2} .
\label{eq:precision}
\end{align}
After the release, however, the field rolls down within a time of order $\tilde m^{-1}\ln(1/\delta_{\rm rel})$, since the Hubble friction is ineffective.

\paragraph{Freeze-out at $\omega_{\rm eff}\simeq H$.}
If the field is to stay near the hilltop after the stabilized phase, the Hubble friction must still be effective at the release, $\tilde m\lesssim H_c$.
In de Sitter space with $\tilde m\lesssim H$, $\omega_{\rm eff}\simeq\tilde m A/A_{\rm th}\propto a^{-3/2}$ decreases relative to $H$,
and the oscillation of $\delta\phi$ becomes overdamped once $\omega_{\rm eff}\simeq H$, i.e.\ when $A\simeq A_{\rm th}H/\tilde m$, well before the end of the stabilized phase.
The subsequent overdamped evolution reduces the deviation only by $\exp(-\int\omega_{\rm eff}^2 dt/3H)$, where the integral is at most $1/9$, and the deviation freezes at
\begin{align}
\delta_* \simeq \delta_i \left(\frac{A_{\rm th}H}{A_i\tilde m}\right)^{1/2} \gtrsim \delta_i\left(\frac{3H}{m_\chi}\right)^{1/2} ,
\label{eq:delta_frozen}
\end{align}
where the second inequality follows from $A_i<A_{\rm max}$.
We have confirmed \eqref{eq:delta_frozen} by integrating the averaged equation for $\delta\phi$ in de Sitter space and taking the envelope over the initial phases:
for $\tilde m/H = 0.3$, $0.1$ and $0.03$ and $A_i/A_{\rm th}=10^3$, we find $\delta_*/\delta_i = 0.061$, $0.099$ and $0.18$, compared with $0.058$, $0.10$ and $0.18$ from \eqref{eq:delta_frozen}.
In a concrete model, such a large ratio $A_i/A_{\rm th}$ requires a large hierarchy: the absence of parametric resonance (see \eqref{eq:noresonance} below), with $e^{3\Delta N}=(A_i/A_{\rm th})^2$, imposes $m_\chi/H\gtrsim 2\times 10^6(\tilde m/H)^2$, i.e.\ $m_\chi/H\gtrsim 2\times 10^5$, $2\times 10^4$ and $2\times 10^3$ for the three cases.
Thus, when the field is to remain frozen after the release, the attractor is much weaker than \eqref{eq:precision} suggests:
the precision is limited by $m_\chi/H$ rather than by $m_\chi/\tilde m$, because the Hubble friction stops the motion before the field reaches the hilltop.
Moreover, once the oscillation is overdamped, quantum fluctuations of order $H/(2\pi f)$ per $e$-fold are no longer damped, which gives an additional floor $\delta_*\gtrsim H/(2\pi f)$.

Finally, the stabilized phase should not be spoiled by the parametric resonance of inhomogeneous modes discussed in \cref{sec:Floquet}.
With $b \to \kappa A/m_\chi$ and $\Omega\to m_\chi$, the condition $b^2\lesssim \Omega H$ reads $\kappa^2 A^2 \lesssim m_\chi^3 H$.
The ratio $\kappa^2A^2/(m_\chi^3 H)$ scales as $a^{-3}$, $a^{-1}$ and $a^{-3/2}$ in de Sitter, radiation and matter domination, respectively,
so that it is sufficient to impose the condition at the beginning of the stabilized phase:
\begin{align}
2\tilde{m}^2 e^{3\Delta N} \lesssim m_\chi H_i .
\label{eq:noresonance}
\end{align}

%%%%%%%%%%%%%%%%%%%%%%%%%%%%%%%%%%%%%%%%%%%%%%%%%%%%%%%%%%%%
\subsection{Decelerating Universe: overdamped stabilization and delayed release}\label{sec:delayed}
%%%%%%%%%%%%%%%%%%%%%%%%%%%%%%%%%%%%%%%%%%%%%%%%%%%%%%%%%%%%

In a decelerating Universe the situation is even more restrictive.
For $A\gg A_{\rm th}$ we have $\omega_{\rm eff}\simeq \tilde m A/A_{\rm th}\propto a^{-3/2}$, while $H\propto a^{-2}$ ($a^{-3/2}$) in radiation (matter) domination,
so that $\omega_{\rm eff}/H$ does not decrease with time, and
\begin{align}
\frac{\omega_{\rm eff}}{H} \lesssim \frac{\tilde m}{H_c}
\label{eq:omega_over_H}
\end{align}
throughout the stabilized phase.
Therefore, if the stabilized phase ends while the Hubble friction is still effective, $H_c\gtrsim\tilde m$, the oscillation is overdamped at all times, and
$\int \omega_{\rm eff}^2\, dt/3H \le \tilde m^2/(4H_c^2)$ in radiation domination:
the attractor does \emph{not} reduce the deviation appreciably.
We have confirmed this numerically for the model \eqref{eq:clean} in radiation domination:
for $H_c/\tilde m = 3$, $2$ and $1$ (starting from $H_i\simeq 12\,\tilde m$), the deviation at the end of the stabilized phase is $0.98$, $0.97$ and $0.79$ times the initial one,
in agreement with $\exp(-\int \omega_{\rm eff}^2 dt/3H)= 0.99$, $0.97$ and $0.86$ (the last one at the $10\%$ level, as expected since $\omega_{\rm eff}/H\simeq 0.7$ there).
In radiation and matter domination, the attractor is effective only if the field is released after the Hubble friction has become ineffective, $H_c\ll\tilde m$, in which case it rolls down soon after the release.

\begin{figure}[t]
\centering
		\includegraphics[width=\columnwidth]{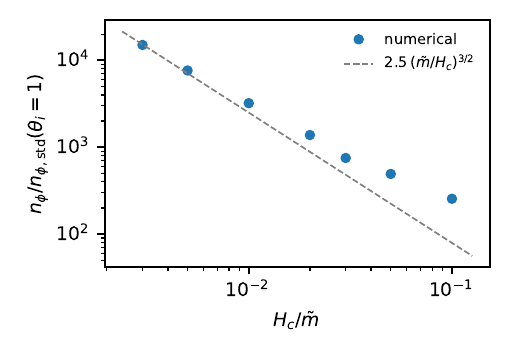}
	\caption{Late-time number density of $\phi$ after the delayed release in radiation domination, normalized by that of the standard misalignment with $\theta_i=1$,
	as a function of the Hubble parameter $H_c$ at the release.
	We use the model \eqref{eq:clean} with $V_0=1$, $\kappa=1.5\times 10^4$, $m_\chi=3\times 10^4$ ($\tilde m^2=0.75$), starting from $H=3$ with $\phi=\pi-0.1$, so that \eqref{eq:noresonance} is satisfied for all points.
	The dashed line shows the scaling \eqref{eq:delayed}.}
	\label{fig:Delayed}
\end{figure}

In a decelerating Universe, the stabilization is instead naturally used to \emph{delay} the onset of the oscillation of $\phi$.
If the field is held at the hilltop until $H_c\ll\tilde m$, the energy density $2m_\phi^2 f^2$ at the hilltop does not redshift until the release,
and the late-time number density of $\phi$ is enhanced, compared with the standard misalignment, by
\begin{align}
\frac{n_\phi}{n_{\phi,\rm std}} \propto \left(\frac{\tilde m}{H_c}\right)^{3/2}
\label{eq:delayed}
\end{align}
in radiation domination, for $H_c\ll\tilde m$.
The leading correction comes from the time $\Delta t\simeq\tilde m^{-1}\ln(1/\delta_{\rm rel})$ needed to roll down from the residual deviation, which delays the onset of the oscillation to $t_c+\Delta t$ and multiplies \eqref{eq:delayed} by $[1+2H_c\ln(1/\delta_{\rm rel})/\tilde m]^{3/2}$; this correction is suppressed by $H_c/\tilde m$.
This is confirmed numerically in \cref{fig:Delayed}:
$n_\phi/n_{\phi,{\rm std}}(\theta_i=1)$ ranges from $2.5\times 10^2$ for $H_c/\tilde m = 0.1$ to $1.5\times 10^4$ for $H_c/\tilde m = 0.003$,
and $(n_\phi/n_{\phi,{\rm std}})(H_c/\tilde m)^{3/2}$ decreases from $8.1$ to $2.5$ over this range, consistently with the above correction; the asymptotic value $\simeq 2.5$ is of the order of the ratio $\simeq 4/\theta_i^2$ of the energy densities at the onset of the oscillation.
Here $n_{\phi,{\rm std}}$ is computed in the same model without the oscillation of $\chi$, where $\phi$ oscillates around $\phi=0$ with mass $(m_\phi^2+\kappa^2/m_\chi^2)^{1/2}$.

This enhancement is phenomenologically similar to that from the late sign flip of the potential in \cite{Murai:2025wbg}, and to the trapped misalignment mechanisms \cite{Nakagawa:2020zjr,DiLuzio:2021gos}, in which the onset of the oscillation is also delayed.
The difference is in how the field is held: here the hilltop remains a maximum of the averaged potential, and the duration of the stabilized phase is set by the Floquet stability window $A_{\rm th}<A<A_{\rm max}$.

Our treatment assumes a homogeneous field.
After a release at $H_c\ll\tilde m$, all modes with $k/a<\tilde m$ are tachyonic, as in the large-misalignment scenario \cite{Arvanitaki:2019rax}.
In the linear regime they grow with rate $(\tilde m^2-k^2/a^2)^{1/2}$, more slowly than the homogeneous mode, so that the relative inhomogeneity does not grow during the roll-down;
however, once the field reaches the anharmonic part of the potential, fragmentation and the formation of compact objects such as oscillons \cite{Gleiser:1993pt,Copeland:1995fq,Kasuya:2002zs,Amin:2011hj} can occur, which our homogeneous treatment does not capture.

%%%%%%%%%%%%%%%%%%%%%%%%%%%%%%%%%%%%%%%%%%%%%%%%%%%%%%%%%%%%
\subsection{General classification}\label{sec:classification}
%%%%%%%%%%%%%%%%%%%%%%%%%%%%%%%%%%%%%%%%%%%%%%%%%%%%%%%%%%%%

The results above can be organized in terms of a few scaling exponents.
Let the oscillating coupling in the Hill equation \eqref{eq:Hill} have amplitude $B_1\propto a^{-\alpha}$ and frequency $\Omega\propto a^{-\beta}$, and let $H\propto a^{-\gamma}$ ($\gamma=0$, $2$, $3/2$ for de Sitter, radiation and matter domination).
Deep in the stabilized phase, the effective frequency of $\delta\phi$ is $\omega_{\rm eff}\simeq B_1/(\sqrt2\,\Omega)\propto a^{-\sigma}$ with
\begin{align}
\sigma\equiv\alpha-\beta .
\end{align}
The following statements then hold for any background:
\begin{enumerate}
\item The Kapitza ratio $B_1^2/(2\Omega^2\tilde m^2)\propto a^{-2\sigma}$ decreases for $\sigma>0$, while the Mathieu parameter $\mathsf{q}=2B_1/\Omega^2\propto a^{2\beta-\alpha}$ decreases for $\alpha\ge 2\beta$.
If both decrease, the stabilized phase ends at the Kapitza threshold, where the hilltop smoothly becomes unstable.
If $\mathsf{q}$ increases (which is always the case for $\sigma<0$ and $\beta>0$), the phase can instead end at the upper edge $\mathsf{q}\simeq 0.91$ of the stability region, where $\delta\phi$ enters an instability band with a growth rate of order $\Omega$, and the field is violently ejected.
\item As long as $\omega_{\rm eff}\gtrsim H$, the adiabatic invariant gives $|\delta\phi|\propto a^{-3/2}\omega_{\rm eff}^{-1/2}\propto a^{-(3-\sigma)/2}$: the hilltop is an attractor for $\sigma<3$, neutral for $\sigma=3$, and repulsive for $\sigma>3$.
\item Since $\omega_{\rm eff}/H\propto a^{\gamma-\sigma}$, the attractor can operate early and then switch off with the field frozen by the Hubble friction ($\tilde m\lesssim H$ at the end) only for $\sigma>\gamma$.
For $\sigma\le\gamma$ we instead have $\omega_{\rm eff}/H\lesssim\tilde m/H_c$ throughout, which generalizes \eqref{eq:omega_over_H}.
\end{enumerate}
For $\gamma<\sigma<3$, the deviation freezes when $\omega_{\rm eff}\simeq H$, at
\begin{align}
\frac{\delta_*}{\delta_i}\simeq\left(\frac{\omega_{{\rm eff},i}}{H_i}\right)^{-\frac{3-\sigma}{2(\sigma-\gamma)}} ,
\label{eq:general_freeze}
\end{align}
which reduces to \eqref{eq:delta_frozen} for $(\sigma,\gamma)=(3/2,0)$.
We have checked \eqref{eq:general_freeze} by integrating the averaged equation for $\delta\phi$ for $(\sigma,\gamma)=(3/2,0)$, $(5/2,0)$, $(5/2,3/2)$, $(5/2,2)$ and $(11/4,2)$ with $\omega_{{\rm eff},i}/H_i=10$--$100$: the dependence on $\omega_{{\rm eff},i}/H_i$ is reproduced, and the prefactor agrees within $25\%$.

\begin{table*}[t]
\caption{Classification of the stabilizers discussed in this paper by the scaling exponent $\sigma$ of $\omega_{\rm eff}\propto a^{-\sigma}$.
``Attract and freeze'' indicates whether the field can be brought towards the hilltop and left frozen by the Hubble friction after the stabilized phase in radiation (RD) or matter (MD) domination.}
\label{tab:classification}
\begin{ruledtabular}
\begin{tabular}{lcccc}
Stabilizer & $\sigma$ & Attraction & End of phase & Attract and freeze \\
\hline
massive field, \eqref{eq:clean} & $3/2$ & $|\delta\phi|\propto a^{-3/4}$ & smooth & no (dS only) \\
spinning axion, \eqref{eq:CoupledAxions} & $-3$ & $|\delta\phi|\propto a^{-3}$ & ejection & no \\
massive field, $m_\chi\propto a^{\nu}$ & $\frac{3}{2}(1+\nu)$ & for $\nu<1$ & smooth & yes for $\frac13<\nu<1$ (RD), $0<\nu<1$ (MD)\\
sleeping top, \cref{sec:top} & $3$ & neutral & smooth & no \\
first-order sign flip \cite{Murai:2025wbg} & -- & -- & -- & no ($\omega_{\rm eff}\lesssim 0.64\,m$) \\
\end{tabular}
\end{ruledtabular}
\end{table*}

The models of this paper are classified in \cref{tab:classification}.
For an oscillating massive field with constant mass, $B_1\propto A\propto a^{-3/2}$ and $\beta=0$, so that $\sigma=3/2$: the attractor works in de Sitter space, but not in radiation or matter domination while the Hubble friction is effective.
For the spinning axion, $B_1$ is constant and $\Omega\propto a^{-3}$, so that $\sigma=-3$: the attractor is strong, but the stabilized phase ends with the ejection of the field, as seen in \cref{TwoAxionTE,TwoAxionSH}.
The sleeping top is not described by the Hill equation, but the restoring frequency in the Routhian \eqref{eq:top_stability} is proportional to $p_\psi\propto a^{-3}$, so that $\sigma=3$ in the same sense: the top is held but not attracted, as seen in \cref{fig:SpinningTop}.
For the first-order sign flip of the averaged potential discussed in \cite{Murai:2025wbg}, the curvature of the averaged potential is that of the bare potential multiplied by a Bessel function $J_0$, so that $\omega_{\rm eff}\le |{\min}J_0|^{1/2}m\simeq 0.64\,m$, where $m$ is the bare mass.
The oscillation around the hilltop is therefore overdamped whenever the Hubble friction is effective, and the deviation is reduced only by $\exp(-\int\omega_{\rm eff}^2dt/3H)$, which in radiation domination is at most $\exp[-0.4\,m^2/(12H^2)]$ by the time the Hubble parameter has decreased to $H$.
The field can approach the hilltop once $H\lesssim 0.64\,m$, as discussed in \cite{Murai:2025wbg}, but then it is no longer frozen by the Hubble friction: there is attraction, but no attract-and-freeze.
The only loophole we find is a massive stabilizer whose mass grows in time, $m_\chi\propto a^{\nu}$: then $A\propto a^{-3/2}m_\chi^{-1/2}$ by the adiabatic invariance, $\sigma=\frac32(1+\nu)$, and the window $\gamma<\sigma<3$ is open for $1/3<\nu<1$ in radiation domination and $0<\nu<1$ in matter domination.
Here we have assumed that the coupling $\kappa$ is approximately constant and that the induced variation of $\tilde m^2=m_\phi^2-\kappa^2/m_\chi^2$ is negligible over the relevant interval; a concrete realization must verify these assumptions as well as adiabaticity, $|\dot m_\chi|\ll m_\chi^2$.
We are not aware of a natural realization of such a mild growth of the mass (the QCD-like temperature dependence $m_\chi\propto T^{-4}\propto a^4$ gives $\sigma=15/2>3$), and the attraction \eqref{eq:general_freeze} is in any case slow, e.g.\ $\delta_*/\delta_i\simeq(\omega_{{\rm eff},i}/H_i)^{-1/2}$ for $\nu=2/3$ in radiation domination.

%%%%%%%%%%%%%%%%%%%%%%%%%%%%%%%%%%%%%%%%%%%%%%%%%%%%%%%%%%%%
\subsection{Energy budget}\label{sec:energy}
%%%%%%%%%%%%%%%%%%%%%%%%%%%%%%%%%%%%%%%%%%%%%%%%%%%%%%%%%%%%

The oscillating field must not dominate the energy density of the Universe, $\rho_\chi\lesssim 3M_P^2H^2$, where $M_P\simeq 2.4\times 10^{18}$ GeV is the reduced Planck mass.
Since $\rho_\chi\propto a^{-3}$, this condition is most stringent at the beginning of the stabilized phase in de Sitter space,
but at the release in radiation domination, where $\rho_\chi/\rho_{\rm rad}\propto a$.

For the attractor in de Sitter space with $\tilde m\lesssim H$, using $\rho_{\chi,i}=\rho_{\rm th}(A_i/A_{\rm th})^2$, \eqref{eq:rho_th}, \eqref{eq:delta_frozen} and $A_i<A_{\rm max}$,
the deviation \eqref{eq:delta_frozen} decreases with $m_\chi$ through $A_{\rm max}$ but the energy density $\rho_{\chi,i}\propto m_\chi^2$ increases.
Minimizing over $m_\chi$, the optimum is $m_\chi\simeq[5\,r\,M_PH\tilde m/(f\sqrt{1-r^2})]^{1/2}$, and the attainable deviation is bounded from below by
\begin{align}
\delta_* \gtrsim \delta_{\rm min} = \left[ \frac{3(1-r^2)}{r^2}\left(\frac{f\,H}{M_P \tilde m}\right)^2\right]^{1/8} \delta_i .
\label{eq:delta_min}
\end{align}
For instance, for $r=1/2$, $f=10^{12}$ GeV and $\tilde m/H=10^{-3}$ we find $\delta_{\rm min}\simeq 0.2\,\delta_i$.
The attractor can therefore bring a field with $\tilde m\ll H$ close to its hilltop only if $f\ll M_P\tilde m/H$.

As a benchmark for the delayed release in radiation domination, consider an axion-like particle with constant mass, held at the hilltop from $H_i=\tilde m$ until $H_c=\tilde m/100$, i.e.\ for $\Delta N\simeq 2.3$ with $A_i/A_{\rm th}\simeq 32$.
The upper bound $A<A_{\rm max}$ requires $m_\chi\gtrsim 10^2\,\tilde m$, and the absence of parametric resonance \eqref{eq:noresonance} requires $m_\chi\gtrsim 2\times 10^3\,\tilde m$; we take $m_\chi=10^4\,\tilde m$ and $r=1/2$.
The fraction of the energy density in $\chi$ at the release is $\rho_{\rm th}/(3M_P^2H_c^2)= (f m_\chi/M_P H_c)^2$, so that $\chi$ remains subdominant for $f\lesssim M_PH_c/m_\chi\simeq 2.4\times 10^{12}$ GeV;
for $f=10^{11}$ GeV the fraction is $2\times 10^{-3}$, and the abundance of $\phi$ is enhanced by $\sim 3\times 10^3$ compared with $\theta_i=1$.

We can estimate the energy transferred to inhomogeneous modes through the narrow first resonance band using the single-crossing production formula \eqref{eq:resonance}.
This estimate assumes an initially unoccupied band, slowly varying background quantities during each crossing, and negligible rescattering and depletion of the stabilizer.
The physical momenta cross the first resonance band at $p=k/a\simeq m_\chi/2$ with $|\dot p|=Hp$.
The phase-space flux through this shell gives
\begin{align}
\dot\rho_{\delta\phi}+4H\rho_{\delta\phi}
&\simeq \frac{H m_\chi^4}{32\pi^2}
\left(e^{\pi R/2}-1\right),
& R&\equiv\frac{\kappa^2A^2}{m_\chi^3H},
\label{eq:rho_fluc_source}
\end{align}
where the produced modes are relativistic because $m_\chi\gg\tilde m$.
Using $\rho_\chi=f^2m_\chi^2A^2/2$ and $\rho_\chi\propto a^{-3}$, the fractional backreaction obeys
\begin{align}
\frac{d\epsilon_{\rm br}}{dN}+\epsilon_{\rm br}
&=\frac{m_\chi^2}{16\pi^2f^2A^2}
\left(e^{\pi R/2}-1\right),
& \epsilon_{\rm br}&\equiv\frac{\rho_{\delta\phi}}{\rho_\chi}.
\label{eq:backreaction_fraction}
\end{align}
For the benchmark above, $R$ decreases from $R_i=0.20$ to $R_c=0.02$ between $H_i=\tilde m$ and $H_c=\tilde m/100$.
Numerically integrating \eqref{eq:backreaction_fraction} with $\epsilon_{\rm br}(N_i)=0$ gives the first-band estimate
\begin{align}
\left.\frac{\rho_{\delta\phi}}{\rho_\chi}\right|_{H_c}
\simeq 1.13\times10^{-5}\left(\frac{m_\chi}{f}\right)^2 .
\label{eq:backreaction_benchmark}
\end{align}
Thus energy transfer through the first band is small relative to the stabilizer energy for this benchmark when $f\gtrsim m_\chi$.
This is a phase-space-integrated estimate based on the local production formula, not a direct evolution of the full fluctuation spectrum in the expanding two-field model.
An energy fraction alone does not establish negligible corrections to the light-field dynamics: these also require control of the fluctuation variance and its contribution to the effective force.
The estimate does not include primordial long-wavelength fluctuations, the low-momentum instability at release, or subsequent nonlinear fragmentation.
For a smaller $f/m_\chi$, a longer stabilized phase, or stronger production, the energy transfer must be evaluated together with these additional effects; extrapolation of \eqref{eq:backreaction_fraction} beyond the narrow-band regime is not justified.

The oscillation of $\chi$ itself behaves as matter, and at the release its energy density exceeds that of $\phi$ by $\rho_{\rm th}/(2f^2m_\phi^2)\sim (m_\chi/\tilde m)^2$.
Therefore $\chi$ must decay, e.g.\ into radiation, after the release but before it dominates the energy density, to avoid an early matter-dominated era and the associated entropy production.
Since $\rho_\chi/\rho_{\rm rad}\propto a\propto t^{1/2}$ grows from $2\times10^{-3}$ at the release, this requires $H_c/(2.5\times 10^5)\lesssim\Gamma_\chi\ll H_c$ for $f=10^{11}$ GeV; alternatively, the decay of $\chi$ can itself trigger the end of the stabilized phase.
A similar conclusion, that the oscillating heavy field generically dominates over the light one and should decay, was reached for the first-order mechanism in \cite{Murai:2025wbg}.
The decay $\chi\to\phi\phi$ induced by the coupling $\Lambda_X^3X\cos\phi$ has the rate $\sim r^2m_\chi m_\phi^2/(32\pi f^2)$, which is negligible for $f\gg m_\chi$.

%%%%%%%%%%%%%%%%%%%%%%%%%%%%%%%%%%%%%%%%%%%%%%%%%%%%%%%%%%%%
\section{Cosmological applications}\label{sec:applications}
%%%%%%%%%%%%%%%%%%%%%%%%%%%%%%%%%%%%%%%%%%%%%%%%%%%%%%%%%%%%

The results of \cref{sec:backgrounds} can be summarized as follows (see also \cref{tab:classification}).
First, if the field is to remain near the hilltop after the stabilized phase, the Hubble friction must be effective at the release.
The attractor then operates only in de Sitter space, and only until $\omega_{\rm eff}\simeq H$, so that the attainable deviation is limited by \eqref{eq:delta_frozen} and by the energy budget \eqref{eq:delta_min}.
Second, in radiation or matter domination, the attractor cannot reduce the deviation while the Hubble friction is effective, see \eqref{eq:omega_over_H},
but the stabilization can delay the onset of the oscillation and enhance the abundance as in \eqref{eq:delayed}.
In all cases the stabilized phase lasts for a few $e$-folds for reasonable hierarchies, too short to sustain the hilltop during the whole period of inflation, or from the early Universe until today.
Let us discuss the implications for each scenario.

\paragraph{Axion dark matter.}
For the QCD axion, an initial misalignment angle $\theta_i$ close to $\pi$ enhances the dark matter abundance through the anharmonic effect \cite{Lyth:1991ub,Visinelli:2009zm},
and such initial conditions have been discussed in \cite{Co:2018mho,Takahashi:2019pqf}.
The abundance scales as
\begin{align}
\Omega_a \propto f_a^{7/6}\, \theta_i^2\, F(\theta_i) , \qquad
F(\theta) \simeq \left[ \ln\frac{e}{1-\theta^2/\pi^2}\right]^{7/6} ,
\label{eq:anharmonic}
\end{align}
where we used the form of $F$ given in Eq.~(40) of \cite{Visinelli:2009zm}, which extends the near-hilltop scaling $F\sim [\ln(e/(1-\theta_i/\pi))]^{1.175}$ for $\theta_i>0.9\pi$ obtained in \cite{Lyth:1991ub} (see also \cite{Turner:1985si,Bae:2008ue}); the exponent $7/6$ corresponds to $m_a(T)\propto T^{-4}$.
Setting $\theta_i$ near $\pi$ by the attractor during inflation is not effective for the QCD axion: since $m_a\ll H_{\rm inf}$, the bound \eqref{eq:delta_min} gives $\delta_{\rm min}\gg 1$ for typical parameters (e.g.\ $\delta_{\rm min}/\delta_i\sim 10^4$ for $f_a=10^{11}$ GeV and $H_{\rm inf}=10^{10}$ GeV), i.e.\ there is no attraction at all.
Even for an axion-like particle with $f H_{\rm inf}\ll M_P\tilde m$, the anharmonic enhancement is modest, e.g.\ $F\simeq 3.7$ for $\delta=0.2$.
In this case $\theta_i$ is fixed during inflation, and the usual isocurvature perturbation of hilltop axions, $S_{\rm iso}\propto(\delta\theta/\delta)/\ln(1/\delta)$ \cite{Kobayashi:2013nva}, applies.

In the radiation-dominated era, \eqref{eq:omega_over_H} shows that the attractor cannot bring the axion closer to the hilltop while $H\gtrsim m_a$.
Instead, the axion can be held at the hilltop beyond the usual onset of oscillation, $H\sim m_a$, and released at $H_c\ll m_a$, which enhances the abundance as in \eqref{eq:delayed}, by $\sim 3\times 10^3$ for $H_c=m_a/100$ (see \cref{sec:energy} for a benchmark).
For the QCD axion, the temperature dependence of $m_a(T)$ makes $A_{\rm th}\propto\tilde m$ grow rapidly, which shortens the stabilized phase; a dedicated analysis is left for future work (see \cite{Murai:2025wbg} for a related discussion of the QCD axion).

In the delayed-release regime, the late-time abundance is determined by the energy density $2m_\phi^2f^2$ held at the hilltop, and depends on the residual deviation only through the logarithmic delay of the roll-down.
The released deviation is a sensitive function of the initial deviation: its envelope scales as \eqref{eq:delta_rel}, but the phase $\Theta_c\simeq\int\omega_{\rm eff}\,dt=\mathcal{O}(\tilde m/H_c)$ is shifted by the anharmonicity by an amount of order $(\tilde m/H_c)\delta_i^2$.
We have checked by solving the full nonlinear equations for the model \eqref{eq:clean} in radiation domination ($\tilde m/H_c\simeq 480$) that the ratio $\delta\phi(t_c+2\tau_c)/\delta_i$ agrees with the linearized result within $0.2\%$ ($5\%$) for $\delta_i=0.01$ ($0.05$), but becomes an erratic function of $\delta_i$ for $\delta_i\gtrsim 0.1$.
Nevertheless, the late-time abundance varies by less than $30\%$ when $\delta_i$ is varied from $0.05$ to $0.85$ (we scanned 49 values):
we find $d\ln n_\phi/d\ln\delta_i = -0.02$, $0.01$, $-0.15$ and $-1.1$ for $\delta_i=0.05$, $0.1$, $0.2$ and $0.4$, respectively,
where the derivative is evaluated locally (the results do not change when the finite-difference step is varied from $10^{-3}$ to $10^{-4}$).
The abundance shows mild spikes, by up to about $10\%$, at isolated values of $\delta_i$ where the released deviation passes through zero, so that the roll-down is delayed (cf.\ the logarithmic correction to \eqref{eq:delayed});
these are the analogue of the spikes found in \cite{Murai:2025wbg}, where the field happens to start oscillating near the hilltop after the final sign flip.
In narrow windows around them, of relative width of order $1\%$ in $\delta_i$, the local derivative becomes larger (e.g.\ $\simeq 4$ at $\delta_i=0.8$).
For this parameter point, the isocurvature perturbation $S_{\rm iso}\simeq (d\ln n_\phi/d\ln\delta_i)\,\delta\theta/\delta_i$ with $\delta\theta\simeq H_{\rm inf}/(2\pi f)$ is therefore not enhanced by the closeness to the hilltop, except in these narrow windows, in contrast to scenarios where $\theta_i$ is simply tuned near $\pi$.
This conclusion is benchmark dependent: a general prediction requires the joint dependence of the abundance on the initial displacement, the amplitude and phase of $\chi$, and their inflationary fluctuations.
If $\chi$ is light during inflation, the fluctuation of its amplitude $\delta A/A$ also changes the release time and hence the abundance by $\delta\ln n_\phi\simeq 2\,\delta A/A$ (from $H_c\propto A_i^{-4/3}$ and \eqref{eq:delayed}), which should be taken into account.
These statements also assume a homogeneous field; see the caveat at the end of \cref{sec:delayed}.

\paragraph{Hilltop inflation.}
In hilltop inflation \cite{Boubekeur:2005zm}, including natural inflation near the top of the axion potential \cite{Freese:1990rb,Nomura:2017ehb}, the inflaton starts near a maximum of its potential.
One might hope to set this initial condition by the Kapitza mechanism.
However, the oscillation of $\chi$ carries the energy density $\rho_\chi\ge\rho_{\rm th}\sim f^2m_\chi^2$, which is much larger than the vacuum energy $\sim f^2 m_\phi^2$ of the inflaton since $m_\chi\gg m_\phi$;
the Universe is then dominated by $\chi$ rather than by the inflaton during the stabilized phase.
A consistent realization would require the stabilized phase to take place during an earlier stage of inflation driven by another field, with $\rho_\chi\lesssim 3M_P^2H^2$, which is difficult to reconcile with $f\gtrsim M_P$ required for natural inflation; we do not pursue this possibility here.

\paragraph{Quintessence.}
Quintessence \cite{Ratra:1987rm,Caldwell:1997ii}, and in particular hilltop quintessence \cite{Dutta:2008qn} and the quintessence axion \cite{Frieman:1995pm,Choi:1999xn,Ibe:2018ffn,Yamazaki:2019bwj} (motivated also by the swampland conjectures \cite{Murayama:2018lie,Ibe:2018ffn}), requires the field to sit near the top of the potential with $m_\phi \gtrsim H_0$, which is a fine-tuning problem discussed in \cite{Choi:1999xn,Ibe:2018ffn}; see also \cite{Bhattacharya:2024kxp} for recent cosmological tests of quintessence models motivated by quantum gravity.
Our results show that the Kapitza mechanism does not alleviate this fine-tuning in a simple way.
Keeping the field at the hilltop from Big Bang nucleosynthesis until today would require $\Delta N \simeq \ln(T_{\rm BBN}/T_0) \simeq 22$ and hence $\rho_{\chi,i}/\rho_{\rm th} \simeq e^{66}$;
the attractor is ineffective in the radiation- and matter-dominated eras with $H\gg m_\phi$, \eqref{eq:omega_over_H};
and during inflation the bound \eqref{eq:delta_min} with $\tilde m/H_{\rm inf}\sim H_0/H_{\rm inf}$ is far from useful.

\paragraph{Curvaton.}
In the curvaton scenario \cite{Enqvist:2001zp,Lyth:2001nq,Moroi:2001ct}, the hilltop curvaton \cite{Kawasaki:2008mc,Kawasaki:2011pd,Fukuda:2018haz} similarly requires the curvaton to sit near the top of its potential during inflation.
The same limitations as above apply:
the attainable deviation is bounded by \eqref{eq:delta_frozen} and \eqref{eq:delta_min},
and the fluctuations generated after the oscillation of the curvaton becomes overdamped are not reduced, so that the curvature perturbation is that of a hilltop curvaton with the deviation $\delta_*$.

\paragraph{Swampland conjectures.}
The refined de Sitter swampland conjecture \cite{Garg:2018reu,Ooguri:2018wrx}, requires either $M_P|\nabla V|\ge c\,V$ or $M_P^2\min(\nabla\nabla V)\le -c'V$ with $c,c'=\mathcal{O}(1)$.
(The original form of the conjecture \cite{Obied:2018sgi} was disfavored by the presence of hilltops in the Higgs and QCD potentials \cite{Denef:2018etk,Murayama:2018lie,Choi:2018rze,Hamaguchi:2018vtv}.)
As discussed in \cite{Fukuda:2018haz}, the second condition favors scalar fields near hilltops:
it requires the inflaton potential to be concave, $\eta_V\le -c'$, which leads to an upper bound on the tensor-to-scalar ratio,
and a curvaton whose potential has a negative curvature of order $-0.01$ in units of the Hubble scale, i.e.\ a hilltop region.
The Kapitza mechanism is compatible with this picture.
The stabilized hilltop is not a critical point of the two-field potential $V(\phi,\chi)$: along the trajectory $\chi$ is displaced from its minimum, so that $\partial_\chi V\neq 0$,
and the curvature in the $\phi$ direction averaged over the oscillation remains negative, $\langle\partial_\phi^2V\rangle=-\tilde m^2$.
No metastable de Sitter minimum of the potential is therefore needed, while the field is nevertheless held near the hilltop for a finite number of $e$-folds.
Whether the swampland conjectures should be applied to the instantaneous potential or to the time-averaged effective potential (cf.\ \eqref{eq:Veff} for the forced model), which has a local minimum at the hilltop, is an open question.
See also \cite{Murayama:2018lie,Ibe:2018ffn} for related implications of the swampland conjectures for axions and quintessence.

\paragraph{Relation to preheating.}
The discussion here is very reminiscent of the Mathieu equation in the discussion of preheating after inflation \cite{Traschen:1990sw,Kofman:1994rk,Shtanov:1994ce,Kofman:1997yn}.
In preheating, one is interested in the instability bands at $\mathsf{a}_k>0$, which lead to the explosive production of particles.
Here, by contrast, we use the stability region at $\mathsf{a}<0$,
where the oscillating background removes the tachyonic instability.
Both phenomena coexist in our setup, and the condition \eqref{eq:noresonance} ensures that the former does not spoil the latter.

%%%%%%%%%%%%%%%%%%%%%%%%%%%%%%%%%%%%%%%%%%%%%%%%%%%%%%%%%%%%
\section{Conclusions}\label{sec:conclusions}
%%%%%%%%%%%%%%%%%%%%%%%%%%%%%%%%%%%%%%%%%%%%%%%%%%%%%%%%%%%%

In this paper we asked whether the hilltop of an axion potential can be
dynamically stabilized by a rapidly oscillating coupling, in analogy with the Kapitza pendulum, and whether this can remove the fine-tuning of hilltop initial conditions.
The stability condition is simply derived from the effective potential in the limit of fast oscillation, and more systematically from the Floquet analysis of the Mathieu (Hill) equation for the fluctuations.
The basin of attraction is finite, \eqref{eq:capture_basin}, so linear stabilization should not be interpreted as capture from an arbitrary initial condition.
We discussed dynamical realizations where the oscillation is provided by another field, as well as a gyroscopic analogue in a sigma model on a squashed three-sphere.

In an expanding Universe, the agent of stabilization is itself damped by the Hubble friction, and the stabilized phase lasts for a number of $e$-folds logarithmic in the hierarchy of the scales.
This number of $e$-folds is independent of the cosmological background, although in radiation and matter domination it corresponds to a physical time that is a power of the hierarchy.
The fate of the field is controlled by the exponent $\sigma$ of $\omega_{\rm eff}\propto a^{-\sigma}$ compared with that of the Hubble parameter, as summarized in \cref{tab:classification}.
During the stabilized phase the hilltop is an attractor, but only as long as $\omega_{\rm eff}\gtrsim H$.
If the field is released after the Hubble friction has become ineffective, the deviation is reduced as $\delta\propto (\rho_{\rm th}/\rho_{\chi,i})^{1/4}$, \eqref{eq:delta_rel}, but the field then rolls down soon after the release.
If instead the field is to remain frozen after the release, the attractor is limited by the Hubble friction, \eqref{eq:delta_frozen}, and by the energy budget, \eqref{eq:delta_min};
in radiation and matter domination it does not reduce the deviation at all, \eqref{eq:omega_over_H}.
The only possible exception in our scaling analysis is a stabilizer whose mass grows slowly in time, subject to additional assumptions on the coupling and the induced light-field mass, for which we are not aware of a natural realization.
These limitations imply that the Kapitza mechanism does not solve the fine-tuning problems of hilltop quintessence or of the initial misalignment angle of the QCD axion set during inflation.
In radiation and matter domination the stabilization can instead delay the onset of the oscillation, enhancing the abundance as $(\tilde m/H_c)^{3/2}$, \eqref{eq:delayed},
and in the delayed-release benchmark the first-band estimate gives an energy transfer of $1.13\times10^{-5}(m_\chi/f)^2$ relative to the stabilizer energy, \eqref{eq:backreaction_benchmark}, subject to the assumptions stated in \cref{sec:energy}.
At the separate parameter point studied for isocurvature, we find no enhancement except in narrow windows of the initial deviation, although this conclusion is benchmark dependent and the phase of the oscillation at the release is scrambled by the anharmonicity.
It would be interesting to find models where the oscillation is sustained, e.g.\ by a continuous energy injection or by a time-dependent background such as the inflaton oscillation, so that the stabilized phase can last longer.
It would also be important to study the inhomogeneous modes, including parametric resonance and the nonlinear dynamics such as the formation of oscillons \cite{Gleiser:1993pt,Copeland:1995fq,Kasuya:2002zs,Amin:2011hj,Hong:2017ooe}, by lattice simulations.
We hope to come back to these questions in the future.

%%%%%%%%%%%%%%%%%%%%%%%%%%%%%%%%%%%%%%%%%%%%%%%%%%%%%%%%%%%%%%%%%%%%%%%%%%%
\section*{Code availability}
The numerical codes used in this paper are available at \url{https://github.com/masahitoyamazaki/inverted-pendulum-in-the-sky} and archived on Zenodo \cite{Numasawa:2026code}.

\section*{Acknowledgments}
T.N.\ is supported by MEXT KAKENHI Grant No.\ 23K13094 and 24H00944 and by JST PRESTO Grant No.\ JPMJPR2359.
M.Y.\ is supported in part by the World Premier International Research Center Initiative (WPI), MEXT, Japan; JSPS KAKENHI Grant No.\ 23K25865; and JST, Japan (CREST Grant No.\ JPMJCR26XA, Moonshot R\&D Grant No.\ JPMJMS256E).
The authors were assisted by Anthropic Claude and OpenAI ChatGPT in writing and editing the manuscript, and in numerical simulations.
%%%%%%%%%%%%%%%%%%%%%%%%%%%%%%%%%%%%%%%%%%%%%%%%%%%%%%%%%%%%
%%%%%%%%%%%%%%%%%%%%%%%%%%%%%%%%%%%%%%%%%%%%%%%%%%%%%%%%%%%%

\bibliography{pendulum}

%%%%%%%%%%%%%%%%%%%%%%%%%%%%%%%%%%%%%%%%%%%%%%%%%%%%%%%%%%%%
%%%%%%%%%%%%%%%%%%%%%%%%%%%%%%%%%%%%%%%%%%%%%%%%%%%%%%%%%%%%

\end{document}